\documentclass[reprint,amssymb,twocolumn,superscriptaddress,aps,prx,10pt]{revtex4-2}

\usepackage{latexsym,amsmath,amssymb,bm,euscript}
\usepackage{float}
\usepackage{amsmath}
\usepackage{amssymb}
\usepackage{amsthm}

\usepackage{amsfonts} 
\usepackage[svgnames]{xcolor}
\usepackage{nicematrix}

\usepackage{tabularx}
\newcolumntype{C}{>{\centering\arraybackslash}X}
\newcolumntype{R}{>{\raggedleft\arraybackslash}X}
\newcolumntype{L}{>{\raggedright\arraybackslash}X}

\usepackage{physics} 
\usepackage{siunitx}
\DeclareSIUnit{\atom}{atom}
\usepackage{bbold}
\usepackage[version=4]{mhchem}
\usepackage{graphicx}
\usepackage{xurl}
\usepackage{hyperref}
\hypersetup{
    colorlinks = true,
    linkcolor  = DodgerBlue,
    citecolor  = CornflowerBlue,
    urlcolor   = DarkSlateBlue,
    }

\usepackage{cleveref} 
\crefname{appendix}{Appendix}{Appendices}
\crefname{equation}{Eq.}{Eqs.}
\crefname{figure}{Fig.}{Figs.}
\crefname{table}{Table}{Tables}
\crefname{section}{Section}{Sections}
\crefname{enumi}{case}{cases}
\creflabelformat{appendix}{[#2#1#3]}

\usepackage{silence}
\newcommand{\nc}{\newcommand}

\nc{\webirvsp}{\href{https://github.com/zjwang11/irvsp}{\texttt{IRVSP} }}
\nc{\webirtb }{\href{https://github.com/zjwang11/irvsp}{\texttt{IR2TB} }}
\nc{\webBRdecomp}{\href{http://tm.iphy.ac.cn/UnconvMat.html}{\texttt{BRdecomp} }}
\nc{\webposabr}{\href{https://github.com/zjwang11/UnconvMat/blob/master/src_pos2aBR.tar.gz}{\texttt{POS2ABR} }}

\nc{\red}[1]{{\color{OrangeRed}{#1}}}
\nc{\redit}[1]{{\color{OrangeRed}{\textit{#1}}}}
\nc{\RED}[1]{{\red{\uppercase{#1}}}}
\nc{\green}[1]{{\color{Teal}{#1}}}
\nc{\blue}[1]{{\color{RoyalBlue}{#1}}}

\nc{\djz}[1]{{\color{Crimson}{\;\text{JD:}\;#1}}}

\nc{\addref}{\red{\emph{(add reference here)}} }

\nc{\prlsection}[1]{\textcolor{blue}{\textit{#1}.---}}

\nc{\eq}{=&\;}
\nc{\eqv}{\equiv&\;}

\nc{\cre}[2]{\hat{#1}_{#2}^{\dagger}}
\nc{\ann}[2]{\hat{#1}_{#2}}
\nc{\ua}{\uparrow}
\nc{\da}{\downarrow}

\def\fs{\text{FS}}

\newcommand{\be}[0]{\begin{equation}}
\newcommand{\ee}[0]{\end{equation}}

\newcommand{\up}[0]{\uparrow}
\newcommand{\dn}[0]{\downarrow}

\def\kk{{\bm{k}}}

\def\pp{\bm{p}}

\def\RR{\bm{R}}
\def\rr{\bm{r}}

\ifdefined\pdfminorversion
\fi

\begin{document}

\title{Topologically Obstructed Kondo Hybridization:\\
A Route to Kondo Semimetals}

\author{Yue Liu}
\affiliation{Department of Physics,
University of Science and Technology of China,
Hefei, Anhui 230026, China}

\author{Junze Deng}
\email{junze.deng@aalto.fi}
\affiliation{Department of Applied Physics,
Aalto University School of Science,
FI-00076 Aalto, Finland}

\author{Haoyu Hu}
\email{huhaoyu314@ustc.edu.cn}
\affiliation{Department of Physics,
University of Science and Technology of China,
Hefei, Anhui 230026, China}


\begin{abstract}
The interplay of strong correlations, topology, and symmetry can reshape the low-energy electronic structure of quantum materials and drive the emergence of novel quantum phases.
In heavy-fermion systems, the Kondo effect entangles localized $f$ moments with conduction electrons, giving rise to coherent heavy-electron bands at low temperatures, typically characterized by a Kondo hybridization gap.
Here we show, however, that the topology and symmetry of the conduction-electron Fermi surface can obstruct the opening of the Kondo hybridization gap at certain momenta.
Treating the Fermi surface of a three-dimensional system as an effective two-dimensional manifold, we demonstrate that nontrivial topological invariants defined on this manifold impose a topological obstruction to opening a full hybridization gap, forcing the $fc$ hybridization to vanish at certain momenta on the Fermi surface. 
These hybridization zeros lead to crossings between $c$ and $f$ bands, giving rise to nodal points or lines near the Fermi energy within the Kondo phase. 
Our results establish topological obstructions to Kondo hybridization as a route to Kondo semimetals, revealing a fundamental connection between Fermi-surface topology and the emergence of correlated electronic phases.
\end{abstract}
\maketitle


\prlsection{Introduction}
$f$-electron systems, commonly described by the Kondo lattice model, provide a paradigmatic setting in which strong local correlations reorganize the electronic structure at low energies~\cite{doniachKondoLatticeWeak1977,hewsonKondoProblemHeavy1993,kondoResistanceMinimumDilute1964,schriefferRelationAndersonKondo1966a,coleman2015introduction,auerbachKondoBosonsKondo1986a,RevModPhys.56.755,gegenwart2008quantum,doi:10.1126/science.1191195,oshikawa2000luttinger}. 
In these systems, the Kondo coupling entangles itinerant conduction electrons with localized $f$ moments, producing composite excitations. 
In the conventional picture, coherent $fc$ hybridization in momentum space produces Kondo bands separated by a direct hybridization gap. 
In parallel, topology and symmetry have emerged as central organizing principles for noninteracting band structures~\cite{bradlyn2017topological,cano2018building,bradlynDiracWeylFermions2016a,armitageWeylDiracSemimetals2018a,hasanColloquiumTopologicalInsulators2010a,qiTopologicalInsulatorsSuperconductors2011a,wanTopologicalSemimetalFermiarc2011a,wengWeylSemimetalPhase2015a,cano2021band,doi:10.1126/science.1245085,PhysRevLett.113.027603}. 
The interplay between such single-particle topology and strong correlations has been explored in heavy-fermion systems~\cite{ma2023nodeline,liu2024antiferromagnetic,kirschbaum2026emergent}, including proposals and realizations of topological Kondo insulators~\cite{alexandrovCubicTopologicalKondo2013a,baruselliKondoHolesTopological2014,dzeroTheoryTopologicalKondo2012a,dzeroTopologicalKondoInsulators2010a,fengCompetingTopologicalKondo2013,hagiwaraSurfaceKondoEffect2016,jiangObservationPossibleTopological2013,kimSurfaceHallEffect2013,kimTopologicalSurfaceState2014,legnerTopologicalInvariantsSurface2014,liTwodimensionalFermiSurfaces2014,Xiang2021,luCorrelatedTopologicalInsulators2013,miyazakiMomentumdependentHybridizationGap2012,neupaneSurfaceElectronicStructure2013a,takimotoSmB6PromisingCandidate2011,tanUnconventionalFermiSurface2015,wengTopologicalCrystallineKondo2014,wolgastLowtemperatureSurfaceConduction2013,xuDirectObservationSpin2014a,xuSurfaceBulkElectronic2013a,zhangHybridizationInterIonCorrelation2013,park2016topological,pirie2020imaging,sato2021surface,gupta2023topological} and Weyl--Kondo semimetals~\cite{laiWeylKondoSemimetal2018a,dzsaberGiantSpontaneousHall2021a,dzsaberKondoInsulatorSemimetal2017,xuHeavyWeylFermion2017,guoEvidenceWeylFermions2018,chen2022topological,dzsaber2022control}. 
In the former case, $fc$ hybridization opens a Kondo gap, giving rise to a Kondo insulator with a nontrivial topological invariant~\cite{dzero2016review}. In the latter case, $f$ states emerge near the Fermi energy and inherit the Weyl character of conduction bands located away from the Fermi energy~\cite{laiWeylKondoSemimetal2018a,grefe2020nonsymmorphic}.

A fundamental question that remains largely unexplored is whether the topology and symmetry properties of the conduction-electron Fermi surface~\cite{haldaneBerryCurvatureFermi2004,xiao2010berry} can actively influence correlated phenomena. 
In superconducting systems, related questions have been studied, and the topological properties of the Fermi surface have been shown to enforce nodal superconductivity~\cite{liTopologicalNodalCooper2018,PhysRevB.105.104515}. 
In this work, we focus on heavy-fermion systems and analyze whether, and how, nontrivial Fermi-surface properties can affect the development of the Kondo hybridization gap.

We demonstrate that the topology and symmetry of the conduction-electron Fermi surface can fundamentally obstruct the opening of a full Kondo hybridization gap. Our starting point is the observation that the Fermi surface of a three-dimensional conduction-electron system forms a two-dimensional manifold in momentum space and can therefore be viewed as an effective two-dimensional system. Such an effective system can carry nontrivial topological or symmetry properties, such as a nonzero spin-resolved Chern number. These properties impose constraints on the $fc$ hybridization and can force the hybridization gap to vanish at certain momenta on the Fermi surface. 
Consequently, the Fermi surface cannot be fully gapped even within the Kondo phase. Crossings between $c$ and $f$ states instead form nodes or nodal lines near the Fermi energy, giving rise to a Kondo semimetal phase through the interplay of electronic correlations and Fermi-surface topology. 

We also comment on the key distinction between our mechanism and previously proposed Weyl--Kondo semimetal scenarios~\cite{laiWeylKondoSemimetal2018a}. In our framework, 
node formation arises from the failure of the Kondo hybridization gap to open.
The resulting nodes are therefore formed jointly by $f$ and $c$ states. By contrast, in the Weyl--Kondo scenario~\cite{laiWeylKondoSemimetal2018a}, the nodes are predominantly formed by $f$-fermion states that inherit the nodal structure of conduction bands located away from the Fermi energy.


\prlsection{Kondo lattice model and Kondo hybridization}
We consider dispersive conduction electrons coupled to localized $f$ orbitals subject to a strong Hubbard interaction $U$. In the large-$U$ limit, the $f$ electrons form local moments, and the resulting low-energy physics is captured by a Kondo-lattice Hamiltonian of the form
\begin{align}
    H_{\text{KL}} = & \sum_{\kk,n,s}\epsilon_{\kk,n} \gamma_{\kk,n,s}^\dag \gamma_{\kk,n,s}
    + \sum_{\substack{
    \kk,\kk',\RR \\
    m,m' \\
    \sigma, s, \sigma', s'
    }} \frac{ {J}_{\kk,\kk'}^{\sigma m s, {\sigma'} m' s' } }{N} \nonumber\\ 
    &
     S^f_{\RR,\sigma\sigma'} \gamma^{\dag}_{\kk',m' s'} \gamma_{\kk, m s}e^{i(\kk-\kk') \cdot (\RR + \rr_f)} 
\end{align}
where $N$ is the number of unit cells, and $\gamma_{\kk,n,s}$ denotes the band annihilation operator for a conduction electron at momentum $\kk$ in the $n$-th band with dispersion $\epsilon_{\kk,n}$. 
For simplicity, we assume that the conduction-electron bands are twofold degenerate, with the two states labeled by $s=\pm$. The degeneracy is protected by $\mathcal{PT}$ symmetry or $SU(2)$ spin symmetry, where $\mathcal{P}$ denotes inversion symmetry and $\mathcal{T}$ denotes time-reversal symmetry. 
We consider a single spin-$1/2$ local moment per unit cell, located at $\rr_f$ and described by the operator $ S^f_{\RR,\sigma\sigma'} = f_{\RR,\sigma}^\dag f_{\RR,\sigma'}$. 
The $f$ fermions satisfy the local constraint $
    \sum_{\sigma}f_{\RR,\sigma}^\dag f_{\RR,\sigma}=1$. 
${J}_{\kk,\kk'}^{\sigma m s, \sigma' m' s'}$ denotes the Kondo coupling obtained via a Schrieffer--Wolff transformation~\cite{schriefferRelationAndersonKondo1966a}.
For an Anderson hybridization of the form $\sum_{\kk,ns,\sigma}[V_\kk]_{ns,\sigma}\gamma_{\kk,ns}^\dagger f_{\kk,\sigma}+\mathrm{h.c.}$, the Kondo coupling is $J_{\kk,\kk'}^{\sigma ms,\sigma'm's'}=(4/U)[V_\kk]_{ms,\sigma}^*[V_{\kk'}]_{m's',\sigma'}$ (see also Appendix S1~\cite{SM}).  
We assume that the system preserves time-reversal symmetry, so that a local Kramers doublet of low-energy $f$-electron states remains. At this stage, we do not specify an explicit form of the $fc$ hybridization and kinetic terms, since our conclusions follow purely from the topological and symmetry properties of the system and do not depend on microscopic details.

For a Kondo system with metallic conduction electrons, the Kondo effect develops at low temperature~\cite{giamarchi2003quantum}. 
The resulting Kondo phase
can be characterized by a nonzero expectation value $\sum_{\kk}  \langle f_{\kk,\sigma}^\dag \gamma_{\kk,ms}\rangle / N \ne 0$ 
within the saddle-point approximation~\cite{bickersReviewTechniquesLarge1987,readSolutionCoqblinSchreifferHamiltonian1983,coleman1984new,millisLargeorbitaldegeneracyExpansionLattice1987,coleman2015introduction}. 
In the Kondo phase, the $f$ fermions acquire a nonzero $U(1)$ charge and become effective itinerant degrees of freedom~\cite{nozieres1974fermi,readSolutionCoqblinSchreifferHamiltonian1983,coleman1984new,coleman2015introduction}.
Assuming that the $n$th conduction band, described by $\gamma_{\kk,ns}$, lies near the Fermi energy, the low-energy theory is then described by the following effective Hamiltonian 
\begin{align} 
    \psi_{\kk,n}^\dag  
    \begin{bmatrix}
    \epsilon_{\kk,n} \mathbb{1}_{2\times 2} & -\Delta_{\kk,n} \\
    -\Delta^\dag_{\kk,n} & \lambda_f \mathbb{1}_{2\times 2}
    \end{bmatrix} \psi_{\kk,n}
\end{align} 
where $\psi_{\kk,n} =\begin{bmatrix}
    \gamma_{\kk,n,+}&\gamma_{\kk,n,-}& f_{\kk,\uparrow}&f_{\kk,\downarrow}
    \end{bmatrix} ^T $. 
$\lambda_f$ is the Lagrange multiplier introduced to enforce the filling constraint of the $f$ fermions. 
$\Delta_{\kk,n}$ is a $2\times2$ matrix characterizing the effective $fc$ hybridization, which is proportional to $\langle f^\dag \gamma\rangle $ and is determined self-consistently. 
For the system with $\mathcal{PT}$ or $SU(2)$ spin symmetry, the two singular values of $\Delta_{\kk,n}$ are identical. We use $\Lambda_{\kk,n}$ to denote the common singular value of $\Delta_{\kk,n}$. 
Diagonalizing this Hamiltonian gives Kondo bands with dispersion
\begin{align} 
    E_{\kk, n, \pm}
    = \frac{\epsilon_{\kk,n}+\lambda_f}{2}
    \pm \sqrt{
    \left(\frac{\epsilon_{\kk,n}-\lambda_f}{2}\right)^2 + \Lambda_{\kk,n}^2
    }\, . 
    \end{align} 
We next introduce the effective Fermi surface of the conduction electrons,
$
    \fs_n = \{\kk \mid \epsilon_{\kk,n} = \lambda_f \approx 0\}$. 
Since $\lambda_f$ lies within the energy window set by the Kondo scale and is exponentially small, the effective $\fs_n$ is nearly identical to the original conduction-electron Fermi surface defined by $\Bqty{\kk \vert \epsilon_{\kk,n} = 0}$. We therefore do not distinguish between the two definitions of the Fermi surface in this manuscript.

\begin{figure}
    \centering
    \includegraphics[width=1.0\linewidth]{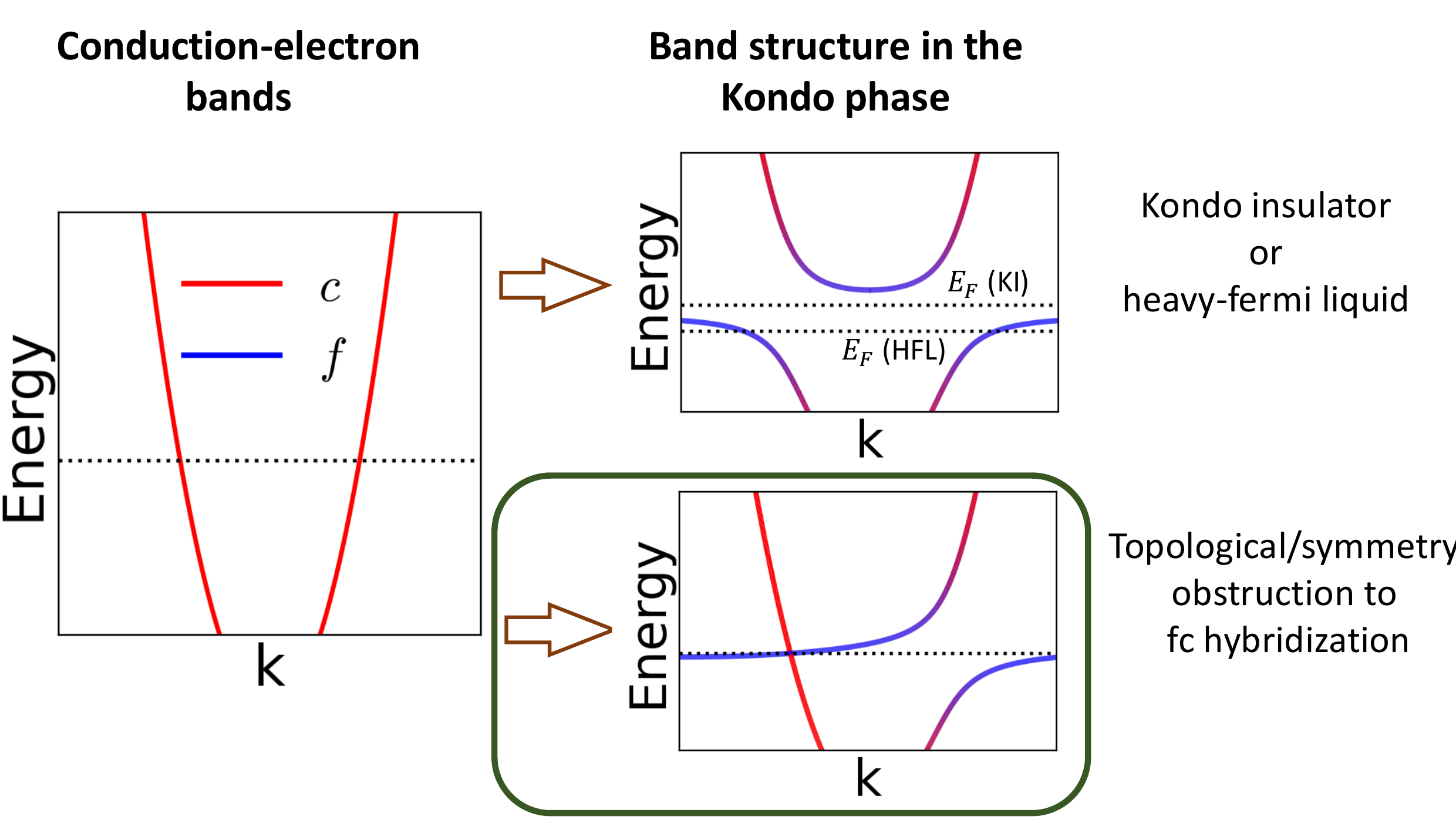}
    \caption{
   Excitation spectra in Kondo systems. In the absence of the Kondo effect, the conduction electrons form a Fermi surface (left). Once Kondo screening develops, a conventional Kondo system exhibits a full $fc$ hybridization gap (top right), giving rise to either a Kondo insulator (KI) or a heavy Fermi liquid (HFL), depending on the position of the Fermi energy $E_F$. By contrast, when the hybridization is obstructed by the topology or symmetry of the conduction-electron Fermi surface, it must vanish at certain momenta. This leads to nodal crossings between the $c$ and $f$ states (bottom right), which is the main focus of this work. 
    }
    \label{fig:illustration}
\end{figure}

In a conventional Kondo system with an on-site Kondo coupling between $c$ and $f$ orbitals, $\Lambda_{\kk,n}$ is nonzero over the entire Fermi surface. 
Consequently, a Kondo hybridization gap opens everywhere on the Fermi surface, with $E_{\kk,n,+}-E_{\kk,n,-}=2\Lambda_{\kk,n}\neq0$ for $\kk\in\fs_n$. This typically leads to a Kondo insulator or to a heavy-fermion metal when the system is doped, as illustrated in~\cref{fig:illustration}.

However, in generic systems, the $c$-electron bands may possess nontrivial topological and symmetry properties. Consequently, the hybridization field $\Lambda_{\kk,n}$ generally acquires a nontrivial momentum dependence inherited from the structure of the conduction-electron wave functions. As a result, the hybridization gap can vanish at certain momenta $\kk^*$ on the Fermi surface, i.e., $\Lambda_{\kk^*,n}=0$. 
As we show below, such zeros can be enforced by the nontrivial topological or symmetry properties of the conduction-electron Fermi surface. 
In this situation, the hybridization gap necessarily closes at those momenta, producing a band crossing between the $c$ and $f$ states near the Fermi energy, characterized by $E_{\kk,n,+}-E_{\kk,n,-}=0$. Consequently, nodes or nodal lines emerge within the Kondo phase, as illustrated in~\cref{fig:illustration}. 
Once the node forms, a weak coupling to additional high-energy bands merely renormalizes the position and velocities of the node without opening a gap~\cite{armitageWeylDiracSemimetals2018a}.

\prlsection{Topological obstructions to the Kondo hybridization} We first consider the case in which the conduction-electron Fermi surface carries a nonzero spin-resolved Chern number, which in turn obstructs the Kondo hybridization. 
Such a topological obstruction can arise in a system with either $SU(2)$ or $U(1)$ spin symmetry, so that the two spin sectors are decoupled, and the hybridization matrix is diagonal,
$\Delta_{\kk,n} = \text{diag}\{ \Delta_{\kk,n,\up},\Delta_{\kk,n,\dn}\}$.  For each spin sector, the Fermi surface can then be treated as an effective two-dimensional system, allowing one to define a spin-resolved Chern number.

The $fc$ hybridization in a given spin sector $\sigma$ enters the Hamiltonian as $-\Delta_{\kk,n,\sigma}\gamma_{\kk,n,\sigma}^\dag f_{\kk,\sigma} + \text{H.c.}$,
with $\Lambda_{\kk,n}=\abs{\Delta_{\kk,n,\sigma}}$. 
Note that both $\Delta_{\kk,n,\sigma}$ and $\gamma_{\kk,n,\sigma}$ possess a $U(1)$ gauge redundancy characterized by $\phi_{\kk,n,\sigma}$,
\begin{equation}
    \gamma_{\kk,n,\sigma}^\dag \rightarrow \gamma_{\kk,n,\sigma}^\dag e^{-i\phi_{\kk,n,\sigma}} \qc
    \Delta_{\kk,n,\sigma}\rightarrow \Delta_{\kk,n,\sigma}e^{i\phi_{\kk,n,\sigma}} .
\end{equation}
A gauge-invariant quantity can be defined as~\cite{liTopologicalNodalCooper2018}
\begin{equation}
    \Phi^\mu_{\kk,\sigma} = \partial_{k^\mu}\text{arg}(\Delta_{\kk,n,\sigma}) - A^\mu_{\kk, n,\sigma} 
\end{equation}
where $A^{\mu}_{\kk, n,\sigma} 
= i\langle \gamma_{\kk,n,\sigma}| \partial_{k_\mu}|\gamma_{\kk,n,\sigma}\rangle 
$ denotes the Berry connection of the conduction-electron band $\gamma_{\kk,n,\sigma}$, and $\partial_{k^\mu}\text{arg}(\Delta_{\kk,n,\sigma}) $ describes the derivative of the phase field of the $fc$ hybridization. $\Phi_{\kk,\sigma}^\mu$ is therefore regular on the Fermi surface, except at nodes of the hybridization where $\Delta_{\kk,n,\sigma} = 0$. 
We use $\mathbf{k}_{i,\sigma}$ to denote the momenta of such hybridization nodes on the Fermi surface. 

\begin{figure}
    \centering
    \includegraphics[width=1.0\linewidth]{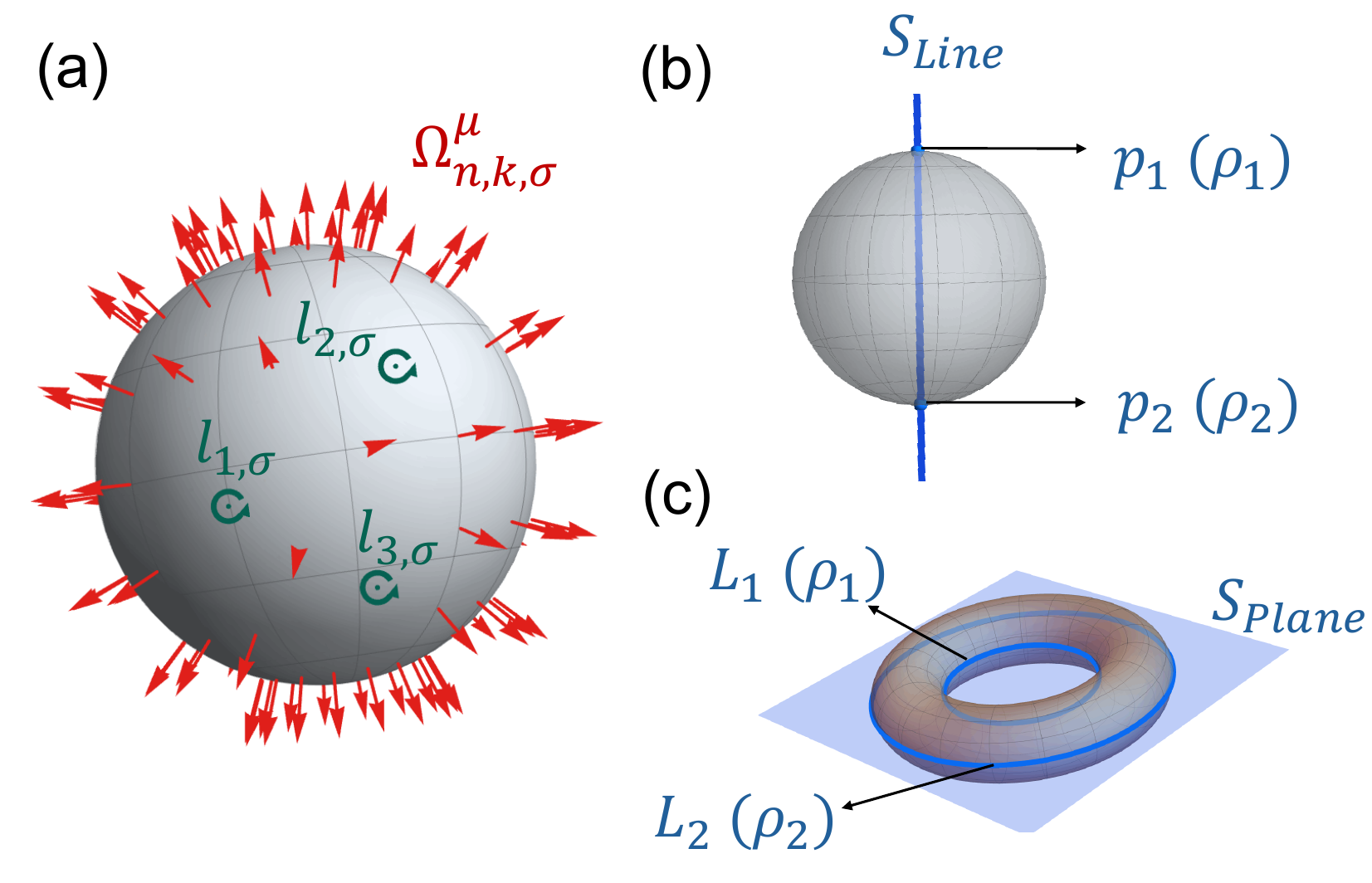}
    \caption{Illustration of the topological and symmetry properties of the Fermi surface.
(a) Fermi surface with nonzero Chern number. The arrows indicate the Berry curvature of electrons on the Fermi surface. The small circles denote infinitesimal loops $\ell_{i,\sigma}$ encircling the nodes of the $fc$ hybridization.
(b) Fermi surface intersecting a high-symmetry line $S_{{Line}}$ of the system. Electrons on the Fermi surface at the intersection points $\pp_{1}$ and $\pp_{2}$ carry irreducible representations $\rho_{1}$ and $\rho_{2}$, respectively. 
(c) Fermi surface intersecting a high-symmetry plane $S_{{Plane}}$ of the system. Electrons on the Fermi surface along the intersection lines $L_{1}$ and $L_{2}$ carry irreducible representations $\rho_{1}$ and $\rho_{2}$, respectively. We define the Fermi surface as nontrivial if $\rho_{1} \neq \rho_{2}$.
    }
    \label{fig:FS_top}
\end{figure}

We then construct an infinitesimal loop $\ell_{i,\sigma}$ on the Fermi surface that encircles an infinitesimal area $S_{i,\sigma}$ containing the node at $\kk_{i,\sigma}$ as illustrated in \cref{fig:FS_top}(a). 
Each such node is characterized by an integer $g_{i,\sigma}$, which corresponds to the vorticity of the phase field $\arg(\Delta_{\kk,n,\sigma})$. 
Explicitly, we have
\begin{equation}
\label{eq:Fermi_surface_winding}
    g_{i,\sigma} =  \oint_{\ell_{i,\sigma}} \sum_\mu \partial_{k^\mu} \arg(\Delta_{\kk,n,\sigma}) \frac{\dd{k_\mu}}{2\pi}
    = \oint_{\ell_{i,\sigma}} \mathbf{\Phi}_{\kk,\sigma} \cdot \frac{\dd{\kk}}{2\pi} 
\end{equation}
Stokes' theorem transforms the sum of loop integrals into an integral over the entire Fermi surface with the small regions surrounding the nodes removed,
\begin{equation}
\label{eq:Fermi_surface_integral}
\begin{aligned}
    &\sum_{i} \oint_{\ell_{i,\sigma}} \vb{\Phi}_{\kk,\sigma} \cdot \frac{\dd{\kk}}{2\pi} 
    = \oint_{\kk \in \fs_n} \frac{\dd{\bm{S}_{\kk}}}{2\pi} \cdot \vb{\Omega}_{\kk,n,\sigma} = \mathcal{C}_{n,\sigma} 
\end{aligned}
\end{equation}
where $\vb{\Omega}_{\kk,n,\sigma}$ is the Berry curvature of the conduction-electron band, and $\mathcal{C}_{n,\sigma}$ corresponds to the Chern number of the Fermi surface $\fs_n$.
Combining \cref{eq:Fermi_surface_winding,eq:Fermi_surface_integral}, we obtain 
\begin{equation}
    \sum_{i} g_{i,\sigma} = \mathcal{C}_{n,\sigma} 
\end{equation} 
We thus conclude that whenever the conduction-electron Fermi surface carries a nonzero Chern number $\mathcal{C}_{n,\sigma} \ne 0$,
$\Delta_{\kk,n,\sigma}$ must have a node on the Fermi surface in order to ensure $\sum_i g_{i,\sigma} \ne 0$.
In other words, there must exist a momentum $\kk_{i,\sigma} \in \fs_n$ such that $\Lambda_{\kk_{i,\sigma},n}
= \abs{\Delta_{\kk_{i,\sigma},n,\sigma}}=0$.
We have therefore established that a nonzero spin-resolved Chern number of the Fermi surface obstructs the $fc$ hybridization and enforces a node.


\prlsection{Symmetry obstructions to Kondo hybridization}
We next consider the case in which the obstruction relies on space-group symmetry. 
Let $S$ denote a high-symmetry submanifold of the Brillouin zone (such as a line or a plane). We associate with $S$ a little group $\mathcal{G}_S$, under which all momenta on $S$ remain invariant~\cite{cano2018building,cano2021band}. 
Suppose the two-dimensional Fermi surface intersects $S$ at momenta $\{\pp_i\}_{i=1,\ldots,n_S}$ when $S$ is a line, or along lines $\{L_i\}_{i=1,\ldots,n_S}$ when $S$ is a plane. 
The Fermi surface can then be viewed as an effective two-dimensional system with symmetry group $\mathcal{G}_S$, featuring high-symmetry points $\{\pp_i\}_i$ or high-symmetry lines $\{L_i\}_i$, as illustrated in \cref{fig:FS_top}(b) and (c), respectively.

We assume that the conduction-electron states on the Fermi surface with momentum $\pp_i$ transform under irreducible representations ${\rho_i}$ of $\mathcal{G}_S$. 
Similarly, for intersections along lines $L_i$, the conduction-electron states at $\kk\in L_i$ transform under irreducible representations $\rho_i$ of $\mathcal{G}_S$.
We call the Fermi surface nontrivial with respect to $\mathcal{G}_S$ if two distinct intersections $i\neq j$ carry different irreducible representations, $\rho_i\neq\rho_j$ (\cref{fig:FS_top}). This representation mismatch can be viewed as a symmetry-protected topological invariant, as it remains unchanged under small symmetry-preserving perturbations of the electronic states.

We now discuss the fate of such a nontrivial Fermi surface when the Kondo effect develops. 
Since the $f$-fermion states originate from a trivial atomic state, they retain the same irreducible representation $\rho_f$ of $\mathcal{G}_S$ throughout the high-symmetry manifold $S$. 
Consequently, for a nontrivial Fermi surface, there must be at least one intersection point $\pp_i$ or intersection line $L_i$ for which $\rho_i\neq\rho_f$. This representation mismatch enforces the vanishing of the Kondo hybridization,
\begin{align}
\rho_i\neq\rho_f  \Rightarrow  \Lambda_{\kk,n}\bigg|_{\kk=\pp_i\text{ or }\kk\in L_i}=0 ,
\end{align}
thereby guaranteeing a band crossing between the $c$ and $f$ states and producing a Dirac node or Dirac nodal line near the Fermi energy.

\begin{figure}
    \centering
    \includegraphics[width=\linewidth]{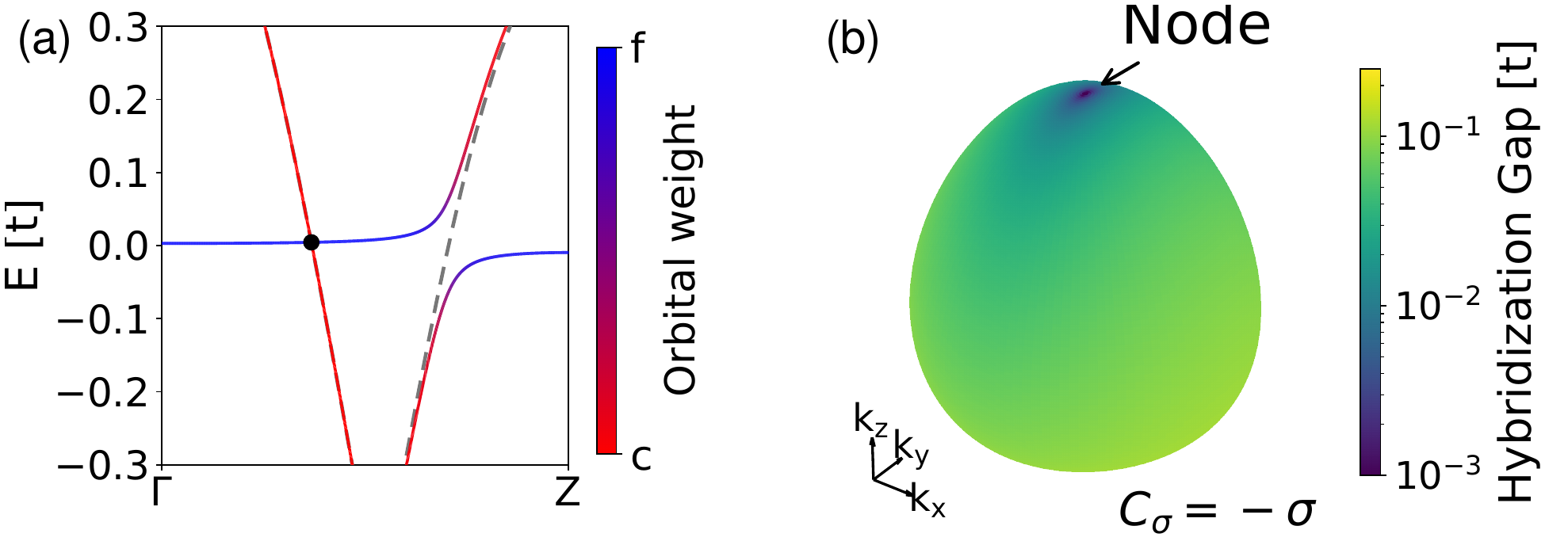}
    \caption{Kondo dispersion and hybridization gap. (a) Dispersion along the path from $\Gamma=(0,0,0)$ to $Z=(0,0,\pi)$. Solid lines show the Kondo bands, with red and blue denoting $c$ and $f$ character, respectively. Gray dashed lines indicate the conduction-electron dispersion in the absence of the Kondo effect. 
    The black dot marks the hybridization node. (b) Hybridization gap $2|\Lambda_{\kk,n}|/t$ on the conduction-electron Fermi surface. The logarithmic color scale is truncated below $10^{-3}$. The arrow marks the node. 
    We use the parameters $t_x=t_y=t_z=t$, $m/t=0.1$, $b/t=0.5$, $\mu/t=0.5$, $V_s=0.35t$, $V_p=0.5t$, $U=10t$, and $\beta t=10^4$.
    }
    \label{fig:model_kondo_dispersion}
\end{figure}

\prlsection{Microscopic model}
We now construct a microscopic model to illustrate the proposed topological obstruction to Kondo hybridization.
A Fermi surface with a nonzero Chern number can be realized by doping Dirac or Weyl nodes. 
We consider a toy model in which the conduction-electron Fermi surface consists of two pockets obtained by doping a three-dimensional Dirac semimetal~\cite{haldaneBerryCurvatureFermi2004,PhysRevB.92.085138,liTopologicalNodalCooper2018}. 
Each unit cell contains two conduction-electron orbitals of $s$ and $p$ character, respectively, both centered at $\rr_c=(0,0,0)$. The localized inversion-odd $f$ orbital is instead centered at $\rr_f=(1/2,1/2,0)$. The Hamiltonian of the conduction electrons can be written as~\cite{wangDiracSemimetalTopological2012,wangThreeDimensionalDirac2013}
\begin{align}
H_c&=\sum_{\kk,\sigma}c_{\kk,\sigma}^\dagger h_\sigma(\kk)c_{\kk,\sigma},\nonumber\\
h_\up(\kk)&=-\mu\tau_0+t_x\sin k_x\,\tau_1
+t_y\sin k_y\,\tau_2+M(\kk)\tau_3,\nonumber\\
h_\dn(\kk)&=h_\up^*(-\kk),\nonumber\\
M(\kk)&=m+t_z\cos k_z+2b(2-\cos k_x-\cos k_y),
\label{eq:main_model_hc}
\end{align}
where $\tau_{1,2,3}$ are Pauli matrices in the orbital basis and $c_{\kk,\sigma}=(c_{\kk,s,\sigma},c_{\kk,p,\sigma})^T$. 
Here $h_\sigma(\kk)$ denotes the renormalized conduction Hamiltonian used in the mean-field calculation. The Kondo coupling in the orbital basis of conduction electrons reads 
{\footnotesize
\begin{align}
J^{\sigma a s,\sigma' b s'}_{\kk,\kk'}
&=\delta_{\sigma,s}\delta_{\sigma',s'}\frac{4}{U}
\left[W_a^\sigma(\kk)\right]^*W_b^{\sigma'}(\kk'),\nonumber\\
W_s^\sigma(\kk)
&=4V_s\left[\sigma\sin\left(\frac{k_x}{2}\right)
\cos\left(\frac{k_y}{2}\right)
+i\cos\left(\frac{k_x}{2}\right)\sin\left(\frac{k_y}{2}\right)\right],\nonumber\\
W_p^\sigma(\kk)
&=4V_p\cos\left(\frac{k_x}{2}\right)
\cos\left(\frac{k_y}{2}\right),
\label{eq:main_model_form_factors}
\end{align}
}
where $V_s$ and $V_p$ characterize the hybridization strengths with the $s$ and $p$ orbitals, respectively.

A self-consistent mean-field calculation yields a Kondo phase at low temperature. The dispersion in this phase is shown in \cref{fig:model_kondo_dispersion}(a). 
A crossing between the $f$ and $c$ bands, marked by the black dot, directly demonstrates the topological obstruction to opening a full Kondo hybridization gap. 
To visualize this obstruction on the Fermi surface, \cref{fig:model_kondo_dispersion}(b) shows the hybridization gap $2|\Delta_{\kk,n,\sigma}|$ on the effective conduction-electron Fermi surface. The hybridization gap is finite everywhere on the Fermi surface except at a single point, revealing the node associated with its nonzero Chern number.
To test the robustness of the node, we also introduce small perturbations that break space-group symmetry. 
The node persists under these perturbations, with only a shift in its momentum-space position (see Appendix S2~\cite{SM} for details).

\prlsection{Summary and discussion}
In this work, we show that the topology and symmetry of the conduction-electron Fermi surface can obstruct the opening of a Kondo hybridization gap, even in the presence of strong local correlations. Treating the Fermi surface as a two-dimensional manifold, we demonstrate that nontrivial topological invariants enforce zeros in the $fc$ hybridization, giving rise to nodal points or lines near the Fermi energy within the Kondo phase. Our results establish topological obstructions to hybridization as a mechanism for realizing Kondo semimetals, demonstrating how Fermi-surface topology constrains the formation of correlated electronic states.


\begin{acknowledgments}
We thank Defa Liu, Yi Jiang, Hanqi Pi, and Lei Chen for helpful discussions.
\end{acknowledgments}

\clearpage
\bibliographystyle{apsrev4-2}
\bibliography{ref}

@article{alexandrovCubicTopologicalKondo2013a,
  author = {Alexandrov, Victor and Dzero, Maxim and Coleman, Piers},
  title = {Cubic topological {Kondo} insulators},
  journal = {Phys. Rev. Lett.},
  volume = {111},
  number = {22},
  pages = {226403},
  year = {2013},
  month = nov,
  doi = {10.1103/PhysRevLett.111.226403},
  issn = {0031-9007, 1079-7114}
}

@article{armitageWeylDiracSemimetals2018a,
  author = {Armitage, N. P. and Mele, E. J. and Vishwanath, Ashvin},
  title = {{Weyl} and {Dirac} semimetals in three-dimensional solids},
  journal = {Rev. Mod. Phys.},
  volume = {90},
  number = {1},
  pages = {015001},
  year = {2018},
  month = jan,
  doi = {10.1103/RevModPhys.90.015001},
  issn = {0034-6861, 1539-0756}
}

@article{baruselliKondoHolesTopological2014,
  author = {Baruselli, Pier Paolo and Vojta, Matthias},
  title = {{Kondo} holes in topological {Kondo} insulators: Spectral properties and surface quasiparticle interference},
  journal = {Phys. Rev. B},
  volume = {89},
  number = {20},
  pages = {205105},
  year = {2014},
  month = may,
  doi = {10.1103/PhysRevB.89.205105},
  issn = {1098-0121, 1550-235X}
}

@article{bradlynDiracWeylFermions2016a,
  author = {Bradlyn, Barry and Cano, Jennifer and Wang, Zhijun and Vergniory, M. G. and Felser, C. and Cava, R. J. and Bernevig, B. Andrei},
  title = {Beyond {Dirac} and {Weyl} fermions: Unconventional quasiparticles in conventional crystals},
  journal = {Science},
  volume = {353},
  number = {6299},
  pages = {aaf5037},
  year = {2016},
  month = aug,
  doi = {10.1126/science.aaf5037},
  issn = {0036-8075, 1095-9203}
}

@article{doniachKondoLatticeWeak1977,
  author = {Doniach, S.},
  title = {The {Kondo} lattice and weak antiferromagnetism},
  journal = {Physica B+C},
  volume = {91},
  pages = {231--234},
  year = {1977},
  month = jul,
  doi = {10.1016/0378-4363(77)90190-5},
  issn = {03784363}
}

@article{dzeroTheoryTopologicalKondo2012a,
  author = {Dzero, Maxim and Sun, Kai and Coleman, Piers and Galitski, Victor},
  title = {Theory of topological {Kondo} insulators},
  journal = {Phys. Rev. B},
  volume = {85},
  number = {4},
  pages = {045130},
  year = {2012},
  month = jan,
  doi = {10.1103/PhysRevB.85.045130},
  issn = {1098-0121, 1550-235X}
}

@article{dzeroTopologicalKondoInsulators2010a,
  author = {Dzero, Maxim and Sun, Kai and Galitski, Victor and Coleman, Piers},
  title = {Topological {Kondo} insulators},
  journal = {Phys. Rev. Lett.},
  volume = {104},
  number = {10},
  pages = {106408},
  year = {2010},
  month = mar,
  doi = {10.1103/PhysRevLett.104.106408},
  issn = {0031-9007, 1079-7114}
}

@article{dzsaberGiantSpontaneousHall2021a,
  author = {Dzsaber, Sami and Yan, Xinlin and Taupin, Mathieu and Eguchi, Gaku and Prokofiev, Andrey and Shiroka, Toni and Blaha, Peter and Rubel, Oleg and Grefe, Sarah E. and Lai, Hsin-Hua and Si, Qimiao and Paschen, Silke},
  title = {Giant spontaneous {Hall} effect in a nonmagnetic {Weyl}--{Kondo} semimetal},
  journal = {Proc. Natl. Acad. Sci. U.S.A.},
  volume = {118},
  number = {8},
  pages = {e2013386118},
  year = {2021},
  month = feb,
  doi = {10.1073/pnas.2013386118},
  issn = {0027-8424, 1091-6490}
}

@article{dzsaberKondoInsulatorSemimetal2017,
  author = {Dzsaber, S. and Prochaska, L. and Sidorenko, A. and Eguchi, G. and Svagera, R. and Waas, M. and Prokofiev, A. and Si, Q. and Paschen, S.},
  title = {{Kondo} insulator to semimetal transformation tuned by spin-orbit coupling},
  journal = {Phys. Rev. Lett.},
  volume = {118},
  number = {24},
  pages = {246601},
  year = {2017},
  month = jun,
  doi = {10.1103/PhysRevLett.118.246601},
  issn = {0031-9007, 1079-7114}
}

@article{fengCompetingTopologicalKondo2013,
  author = {Feng, Xiao-Yong and Dai, Jianhui and Chung, Chung-Hou and Si, Qimiao},
  title = {Competing topological and {Kondo} insulator phases on a honeycomb lattice},
  journal = {Phys. Rev. Lett.},
  volume = {111},
  number = {1},
  pages = {016402},
  year = {2013},
  month = jul,
  doi = {10.1103/PhysRevLett.111.016402},
  issn = {0031-9007, 1079-7114}
}

@article{guoEvidenceWeylFermions2018,
  author = {Guo, C. Y. and Wu, F. and Wu, Z. Z. and Smidman, M. and Cao, C. and Bostwick, A. and Jozwiak, C. and Rotenberg, E. and Liu, Y. and Steglich, F. and Yuan, H. Q.},
  title = {Evidence for {Weyl} fermions in a canonical heavy-fermion semimetal {YbPtBi}},
  journal = {Nat. Commun.},
  volume = {9},
  number = {1},
  pages = {4622},
  year = {2018},
  month = nov,
  doi = {10.1038/s41467-018-06782-1},
  issn = {2041-1723}
}

@article{hagiwaraSurfaceKondoEffect2016,
  author = {Hagiwara, Kenta and Ohtsubo, Yoshiyuki and Matsunami, Masaharu and Ideta, Shin-ichiro and Tanaka, Kiyohisa and Miyazaki, Hidetoshi and Rault, Julien E. and {Le F{\`e}vre}, Patrick and Bertran, Fran{\c c}ois and {Taleb-Ibrahimi}, Amina and Yukawa, Ryu and Kobayashi, Masaki and Horiba, Koji and Kumigashira, Hiroshi and Sumida, Kazuki and Okuda, Taichi and Iga, Fumitoshi and Kimura, Shin-ichi},
  title = {Surface {Kondo} effect and non-trivial metallic state of the {Kondo} insulator {YbB$_{12}$}},
  journal = {Nat. Commun.},
  volume = {7},
  number = {1},
  pages = {12690},
  year = {2016},
  month = aug,
  doi = {10.1038/ncomms12690},
  issn = {2041-1723}
}

@article{haldaneBerryCurvatureFermi2004,
  author = {Haldane, F. D. M.},
  title = {{Berry} curvature on the {Fermi} surface: Anomalous {Hall} effect as a topological {Fermi}-liquid property},
  journal = {Phys. Rev. Lett.},
  volume = {93},
  number = {20},
  pages = {206602},
  year = {2004},
  month = nov,
  doi = {10.1103/PhysRevLett.93.206602},
  issn = {0031-9007, 1079-7114}
}

@article{hasanColloquiumTopologicalInsulators2010a,
  author = {Hasan, M. Z. and Kane, C. L.},
  title = {Colloquium: Topological insulators},
  journal = {Rev. Mod. Phys.},
  volume = {82},
  number = {4},
  pages = {3045--3067},
  year = {2010},
  month = nov,
  doi = {10.1103/RevModPhys.82.3045},
  issn = {0034-6861, 1539-0756}
}

@book{hewsonKondoProblemHeavy1993,
  author = {Hewson, Alexander Cyril},
  title = {The {Kondo} Problem to Heavy Fermions},
  publisher = {Cambridge University Press},
  address = {Cambridge},
  edition = {1},
  year = {1993},
  month = jan,
  doi = {10.1017/CBO9780511470752},
  isbn = {978-0-521-36382-2}
}

@article{jiangObservationPossibleTopological2013,
  author = {Jiang, J. and Li, S. and Zhang, T. and Sun, Z. and Chen, F. and Ye, Z. R. and Xu, M. and Ge, Q. Q. and Tan, S. Y. and Niu, X. H. and Xia, M. and Xie, B. P. and Li, Y. F. and Chen, X. H. and Wen, H. H. and Feng, D. L.},
  title = {Observation of possible topological in-gap surface states in the {Kondo} insulator {SmB$_6$} by photoemission},
  journal = {Nat. Commun.},
  volume = {4},
  number = {1},
  pages = {3010},
  year = {2013},
  month = dec,
  doi = {10.1038/ncomms4010},
  issn = {2041-1723}
}

@article{kimSurfaceHallEffect2013,
  author = {Kim, D. J. and Thomas, S. and Grant, T. and Botimer, J. and Fisk, Z. and Xia, Jing},
  title = {Surface {Hall} effect and nonlocal transport in {SmB$_6$}: Evidence for surface conduction},
  journal = {Sci. Rep.},
  volume = {3},
  number = {1},
  pages = {3150},
  year = {2013},
  month = nov,
  doi = {10.1038/srep03150},
  issn = {2045-2322}
}

@article{kimTopologicalSurfaceState2014,
  author = {Kim, D. J. and Xia, J. and Fisk, Z.},
  title = {Topological surface state in the {Kondo} insulator samarium hexaboride},
  journal = {Nat. Mater.},
  volume = {13},
  number = {5},
  pages = {466--470},
  year = {2014},
  month = may,
  doi = {10.1038/nmat3913},
  issn = {1476-1122, 1476-4660}
}

@article{kondoResistanceMinimumDilute1964,
  author = {Kondo, J.},
  title = {Resistance minimum in dilute magnetic alloys},
  journal = {Prog. Theor. Phys.},
  volume = {32},
  number = {1},
  pages = {37--49},
  year = {1964},
  month = jul,
  doi = {10.1143/PTP.32.37},
  issn = {0033-068X, 1347-4081}
}

@article{laiWeylKondoSemimetal2018a,
  author = {Lai, Hsin-Hua and Grefe, Sarah E. and Paschen, Silke and Si, Qimiao},
  title = {{Weyl}--{Kondo} semimetal in heavy-fermion systems},
  journal = {Proc. Natl. Acad. Sci. U.S.A.},
  volume = {115},
  number = {1},
  pages = {93--97},
  year = {2018},
  month = jan,
  doi = {10.1073/pnas.1715851115},
  issn = {0027-8424, 1091-6490}
}

@article{legnerTopologicalInvariantsSurface2014,
  author = {Legner, Markus and R{\"u}egg, Andreas and Sigrist, Manfred},
  title = {Topological invariants, surface states, and interaction-driven phase transitions in correlated {Kondo} insulators with cubic symmetry},
  journal = {Phys. Rev. B},
  volume = {89},
  number = {8},
  pages = {085110},
  year = {2014},
  month = feb,
  doi = {10.1103/PhysRevB.89.085110},
  issn = {1098-0121, 1550-235X}
}

@article{liTopologicalNodalCooper2018,
  author = {Li, Yi and Haldane, F. D. M.},
  title = {Topological nodal {Cooper} pairing in doped {Weyl} metals},
  journal = {Phys. Rev. Lett.},
  volume = {120},
  number = {6},
  pages = {067003},
  year = {2018},
  month = feb,
  doi = {10.1103/PhysRevLett.120.067003},
  issn = {0031-9007, 1079-7114}
}

@article{liTwodimensionalFermiSurfaces2014,
  author = {Li, G. and Xiang, Z. and Yu, F. and Asaba, T. and Lawson, B. and Cai, P. and Tinsman, C. and Berkley, A. and Wolgast, S. and Eo, Y. S. and Kim, Dae-Jeong and Kurdak, C. and Allen, J. W. and Sun, K. and Chen, X. H. and Wang, Y. Y. and Fisk, Z. and Li, Lu},
  title = {Two-dimensional {Fermi} surfaces in {Kondo} insulator {SmB$_6$}},
  journal = {Science},
  volume = {346},
  number = {6214},
  pages = {1208--1212},
  year = {2014},
  month = dec,
  doi = {10.1126/science.1250366},
  issn = {0036-8075, 1095-9203}
}

@article{luCorrelatedTopologicalInsulators2013,
  author = {Lu, Feng and Zhao, JianZhou and Weng, Hongming and Fang, Zhong and Dai, Xi},
  title = {Correlated topological insulators with mixed valence},
  journal = {Phys. Rev. Lett.},
  volume = {110},
  number = {9},
  pages = {096401},
  year = {2013},
  month = feb,
  doi = {10.1103/PhysRevLett.110.096401},
  issn = {0031-9007, 1079-7114}
}

@article{miyazakiMomentumdependentHybridizationGap2012,
  author = {Miyazaki, Hidetoshi and Hajiri, Tetsuya and Ito, Takahiro and Kunii, Satoru and Kimura, Shin-ichi},
  title = {Momentum-dependent hybridization gap and dispersive in-gap state of the {Kondo} semiconductor {SmB$_6$}},
  journal = {Phys. Rev. B},
  volume = {86},
  number = {7},
  pages = {075105},
  year = {2012},
  month = aug,
  doi = {10.1103/PhysRevB.86.075105},
  issn = {1098-0121, 1550-235X}
}

@article{neupaneSurfaceElectronicStructure2013a,
  author = {Neupane, M. and Alidoust, N. and Xu, S.-Y. and Kondo, T. and Ishida, Y. and Kim, D. J. and Liu, Chang and Belopolski, I. and Jo, Y. J. and Chang, T.-R. and Jeng, H.-T. and Durakiewicz, T. and Balicas, L. and Lin, H. and Bansil, A. and Shin, S. and Fisk, Z. and Hasan, M. Z.},
  title = {Surface electronic structure of the topological {Kondo}-insulator candidate correlated electron system {SmB$_6$}},
  journal = {Nat. Commun.},
  volume = {4},
  number = {1},
  pages = {2991},
  year = {2013},
  month = dec,
  doi = {10.1038/ncomms3991},
  issn = {2041-1723}
}

@article{qiTopologicalInsulatorsSuperconductors2011a,
  author = {Qi, Xiao-Liang and Zhang, Shou-Cheng},
  title = {Topological insulators and superconductors},
  journal = {Rev. Mod. Phys.},
  volume = {83},
  number = {4},
  pages = {1057--1110},
  year = {2011},
  month = oct,
  doi = {10.1103/RevModPhys.83.1057},
  issn = {0034-6861, 1539-0756}
}

@article{schriefferRelationAndersonKondo1966a,
  author = {Schrieffer, J. R. and Wolff, P. A.},
  title = {Relation between the {Anderson} and {Kondo} {Hamiltonians}},
  journal = {Phys. Rev.},
  volume = {149},
  number = {2},
  pages = {491--492},
  year = {1966},
  month = sep,
  doi = {10.1103/PhysRev.149.491},
  issn = {0031-899X}
}

@article{takimotoSmB6PromisingCandidate2011,
  author = {Takimoto, Tetsuya},
  title = {{SmB$_6$}: A promising candidate for a topological insulator},
  journal = {J. Phys. Soc. Jpn.},
  volume = {80},
  number = {12},
  pages = {123710},
  year = {2011},
  month = dec,
  doi = {10.1143/JPSJ.80.123710},
  issn = {0031-9015, 1347-4073}
}

@article{tanUnconventionalFermiSurface2015,
  author = {Tan, B. S. and Hsu, Y.-T. and Zeng, B. and {Ciomaga Hatnean}, Monica and Harrison, N. and Zhu, Z. and Hartstein, M. and Kiourlappou, M. and Srivastava, A. and Johannes, M. D. and Murphy, T. P. and Park, J.-H. and Balicas, L. and Lonzarich, G. G. and Balakrishnan, G. and Sebastian, Suchitra E.},
  title = {Unconventional {Fermi} surface in an insulating state},
  journal = {Science},
  volume = {349},
  number = {6245},
  pages = {287--290},
  year = {2015},
  month = jul,
  doi = {10.1126/science.aaa7974},
  issn = {0036-8075, 1095-9203}
}

@article{wanTopologicalSemimetalFermiarc2011a,
  author = {Wan, Xiangang and Turner, Ari M. and Vishwanath, Ashvin and Savrasov, Sergey Y.},
  title = {Topological semimetal and {Fermi}-arc surface states in the electronic structure of pyrochlore iridates},
  journal = {Phys. Rev. B},
  volume = {83},
  number = {20},
  pages = {205101},
  year = {2011},
  month = may,
  doi = {10.1103/PhysRevB.83.205101},
  issn = {1098-0121, 1550-235X}
}

@article{wengTopologicalCrystallineKondo2014,
  author = {Weng, Hongming and Zhao, Jianzhou and Wang, Zhijun and Fang, Zhong and Dai, Xi},
  title = {Topological crystalline {Kondo} insulator in mixed valence ytterbium borides},
  journal = {Phys. Rev. Lett.},
  volume = {112},
  number = {1},
  pages = {016403},
  year = {2014},
  month = jan,
  doi = {10.1103/PhysRevLett.112.016403},
  issn = {0031-9007, 1079-7114}
}

@article{wengWeylSemimetalPhase2015a,
  author = {Weng, Hongming and Fang, Chen and Fang, Zhong and Bernevig, B. Andrei and Dai, Xi},
  title = {{Weyl} semimetal phase in noncentrosymmetric transition-metal monophosphides},
  journal = {Phys. Rev. X},
  volume = {5},
  number = {1},
  pages = {011029},
  year = {2015},
  month = mar,
  doi = {10.1103/PhysRevX.5.011029},
  issn = {2160-3308}
}

@article{wolgastLowtemperatureSurfaceConduction2013,
  author = {Wolgast, Steven and Kurdak, {\c C}a{\u g}l{\i}yan and Sun, Kai and Allen, J. W. and Kim, Dae-Jeong and Fisk, Zachary},
  title = {Low-temperature surface conduction in the {Kondo} insulator {SmB$_6$}},
  journal = {Phys. Rev. B},
  volume = {88},
  number = {18},
  pages = {180405(R)},
  year = {2013},
  month = nov,
  doi = {10.1103/PhysRevB.88.180405},
  issn = {1098-0121, 1550-235X}
}

@article{xuDirectObservationSpin2014a,
  author = {Xu, N. and Biswas, P. K. and Dil, J. H. and Dhaka, R. S. and Landolt, G. and Muff, S. and Matt, C. E. and Shi, X. and Plumb, N. C. and Radovi{\'c}, M. and Pomjakushina, E. and Conder, K. and Amato, A. and Borisenko, S. V. and Yu, R. and Weng, H.-M. and Fang, Z. and Dai, X. and Mesot, J. and Ding, H. and Shi, M.},
  title = {Direct observation of the spin texture in {SmB$_6$} as evidence of the topological {Kondo} insulator},
  journal = {Nat. Commun.},
  volume = {5},
  number = {1},
  pages = {4566},
  year = {2014},
  month = jul,
  doi = {10.1038/ncomms5566},
  issn = {2041-1723}
}

@article{xuHeavyWeylFermion2017,
  author = {Xu, Yuanfeng and Yue, Changming and Weng, Hongming and Dai, Xi},
  title = {Heavy {Weyl} fermion state in {CeRu$_4$Sn$_6$}},
  journal = {Phys. Rev. X},
  volume = {7},
  number = {1},
  pages = {011027},
  year = {2017},
  month = mar,
  doi = {10.1103/PhysRevX.7.011027},
  issn = {2160-3308}
}

@article{xuSurfaceBulkElectronic2013a,
  author = {Xu, N. and Shi, X. and Biswas, P. K. and Matt, C. E. and Dhaka, R. S. and Huang, Y. and Plumb, N. C. and Radovi{\'c}, M. and Dil, J. H. and Pomjakushina, E. and Conder, K. and Amato, A. and Salman, Z. and Paul, D. {\relax McK}. and Mesot, J. and Ding, H. and Shi, M.},
  title = {Surface and bulk electronic structure of the strongly correlated system {SmB$_6$} and implications for a topological {Kondo} insulator},
  journal = {Phys. Rev. B},
  volume = {88},
  number = {12},
  pages = {121102(R)},
  year = {2013},
  month = sep,
  doi = {10.1103/PhysRevB.88.121102},
  issn = {1098-0121, 1550-235X}
}

@article{zhangHybridizationInterIonCorrelation2013,
  author = {Zhang, Xiaohang and Butch, N. P. and Syers, P. and Ziemak, S. and Greene, Richard L. and Paglione, Johnpierre},
  title = {Hybridization, inter-ion correlation, and surface states in the {Kondo} insulator {SmB$_6$}},
  journal = {Phys. Rev. X},
  volume = {3},
  number = {1},
  pages = {011011},
  year = {2013},
  month = feb,
  doi = {10.1103/PhysRevX.3.011011},
  issn = {2160-3308}
}

@book{coleman2015introduction,
  author = {Coleman, Piers},
  title = {Introduction to Many-Body Physics},
  publisher = {Cambridge University Press},
  address = {Cambridge},
  year = {2015},
  doi = {10.1017/CBO9781139020916},
  isbn = {978-0-521-86488-6}
}

@article{bradlyn2017topological,
  author = {Bradlyn, Barry and Elcoro, Luis and Cano, Jennifer and Vergniory, Maia G. and Wang, Zhijun and Felser, Claudia and Aroyo, Mois I. and Bernevig, B. Andrei},
  title = {Topological quantum chemistry},
  journal = {Nature},
  publisher = {Nature Publishing Group UK London},
  volume = {547},
  number = {7663},
  pages = {298--305},
  year = {2017},
  doi = {10.1038/nature23268}
}

@article{cano2018building,
  author = {Cano, Jennifer and Bradlyn, Barry and Wang, Zhijun and Elcoro, Luis and Vergniory, Maia Garcia and Felser, Claudia and Aroyo, Mois I. and Bernevig, B. Andrei},
  title = {Building blocks of topological quantum chemistry: Elementary band representations},
  journal = {Phys. Rev. B},
  publisher = {APS},
  volume = {97},
  number = {3},
  pages = {035139},
  year = {2018},
  doi = {10.1103/PhysRevB.97.035139}
}

@article{cano2021band,
  author = {Cano, Jennifer and Bradlyn, Barry},
  title = {Band representations and topological quantum chemistry},
  journal = {Annu. Rev. Condens. Matter Phys.},
  publisher = {Annual Reviews},
  volume = {12},
  number = {1},
  pages = {225--246},
  year = {2021},
  doi = {10.1146/annurev-conmatphys-041720-124134}
}

@article{auerbachKondoBosonsKondo1986a,
  author = {Auerbach, Assa and Levin, K.},
  title = {{Kondo} bosons and the {Kondo} lattice: Microscopic basis for the heavy {Fermi} liquid},
  journal = {Phys. Rev. Lett.},
  volume = {57},
  number = {7},
  pages = {877--880},
  year = {1986},
  month = aug,
  doi = {10.1103/PhysRevLett.57.877},
  issn = {0031-9007}
}

@article{bickersReviewTechniquesLarge1987,
  author = {Bickers, N. E.},
  title = {Review of techniques in the large-{$N$} expansion for dilute magnetic alloys},
  journal = {Rev. Mod. Phys.},
  volume = {59},
  number = {4},
  pages = {845--939},
  year = {1987},
  month = oct,
  doi = {10.1103/RevModPhys.59.845},
  issn = {0034-6861}
}

@article{millisLargeorbitaldegeneracyExpansionLattice1987,
  author = {Millis, A. J. and Lee, P. A.},
  title = {Large-orbital-degeneracy expansion for the lattice {Anderson} model},
  journal = {Phys. Rev. B},
  volume = {35},
  number = {7},
  pages = {3394--3414},
  year = {1987},
  month = mar,
  doi = {10.1103/PhysRevB.35.3394},
  issn = {0163-1829}
}

@article{readSolutionCoqblinSchreifferHamiltonian1983,
  author = {Read, N. and Newns, D. M.},
  title = {On the solution of the {Coqblin--Schrieffer} {Hamiltonian} by the large-{$N$} expansion technique},
  journal = {J. Phys. C: Solid State Phys.},
  volume = {16},
  number = {17},
  pages = {3273--3295},
  year = {1983},
  month = jun,
  doi = {10.1088/0022-3719/16/17/014},
  issn = {0022-3719}
}

@article{coleman1984new,
  author = {Coleman, Piers},
  title = {New approach to the mixed-valence problem},
  journal = {Phys. Rev. B},
  publisher = {APS},
  volume = {29},
  number = {6},
  pages = {3035--3044},
  year = {1984},
  doi = {10.1103/PhysRevB.29.3035}
}

@article{xiao2010berry,
  author = {Xiao, Di and Chang, Ming-Che and Niu, Qian},
  title = {{Berry} phase effects on electronic properties},
  journal = {Rev. Mod. Phys.},
  publisher = {APS},
  volume = {82},
  number = {3},
  pages = {1959--2007},
  year = {2010},
  doi = {10.1103/RevModPhys.82.1959}
}

@article{chen2022topological,
  author = {Chen, Lei and Setty, Chandan and Hu, Haoyu and Vergniory, Maia G. and Grefe, Sarah E. and Fischer, Lukas and Yan, Xinlin and Eguchi, Gaku and Prokofiev, Andrey and Paschen, Silke and Cano, Jennifer and Si, Qimiao},
  title = {Topological semimetal driven by strong correlations and crystalline symmetry},
  journal = {Nat. Phys.},
  publisher = {Nature Publishing Group UK London},
  volume = {18},
  number = {11},
  pages = {1341--1346},
  year = {2022},
  doi = {10.1038/s41567-022-01743-4}
}

@article{doi:10.1126/science.1245085,
  author = {Liu, Z. K. and Zhou, B. and Zhang, Y. and Wang, Z. J. and Weng, H. M. and Prabhakaran, D. and Mo, S.-K. and Shen, Z. X. and Fang, Z. and Dai, X. and Hussain, Z. and Chen, Y. L.},
  title = {Discovery of a three-dimensional topological {Dirac} semimetal, {Na$_3$Bi}},
  journal = {Science},
  volume = {343},
  number = {6173},
  pages = {864--867},
  year = {2014},
  doi = {10.1126/science.1245085}
}

@article{PhysRevLett.113.027603,
  author = {Borisenko, Sergey and Gibson, Quinn and Evtushinsky, Danil and Zabolotnyy, Volodymyr and B\"uchner, Bernd and Cava, Robert J.},
  title = {Experimental realization of a three-dimensional {Dirac} semimetal},
  journal = {Phys. Rev. Lett.},
  publisher = {American Physical Society},
  volume = {113},
  number = {2},
  pages = {027603},
  year = {2014},
  month = jul,
  doi = {10.1103/PhysRevLett.113.027603},
  url = {https://doi.org/10.1103/PhysRevLett.113.027603},
  numpages = {5}
}

@article{RevModPhys.56.755,
  author = {Stewart, G. R.},
  title = {Heavy-fermion systems},
  journal = {Rev. Mod. Phys.},
  publisher = {American Physical Society},
  volume = {56},
  number = {4},
  pages = {755--787},
  year = {1984},
  month = oct,
  doi = {10.1103/RevModPhys.56.755},
  url = {https://doi.org/10.1103/RevModPhys.56.755}
}

@article{gegenwart2008quantum,
  author = {Gegenwart, Philipp and Si, Qimiao and Steglich, Frank},
  title = {Quantum criticality in heavy-fermion metals},
  journal = {Nat. Phys.},
  publisher = {Nature Publishing Group UK London},
  volume = {4},
  number = {3},
  pages = {186--197},
  year = {2008},
  doi = {10.1038/nphys892}
}

@article{doi:10.1126/science.1191195,
  author = {Si, Qimiao and Steglich, Frank},
  title = {Heavy fermions and quantum phase transitions},
  journal = {Science},
  volume = {329},
  number = {5996},
  pages = {1161--1166},
  year = {2010},
  doi = {10.1126/science.1191195},
  url = {https://doi.org/10.1126/science.1191195}
}

@article{nozieres1974fermi,
  author = {Nozi{\`e}res, Philippe},
  title = {A ``{Fermi}-liquid'' description of the {Kondo} problem at low temperatures},
  journal = {J. Low Temp. Phys.},
  publisher = {Springer},
  volume = {17},
  number = {1--2},
  pages = {31--42},
  year = {1974},
  doi = {10.1007/BF00654541}
}

@book{giamarchi2003quantum,
  author = {Giamarchi, Thierry},
  title = {Quantum Physics in One Dimension},
  series = {International Series of Monographs on Physics},
  volume = {121},
  publisher = {Clarendon Press},
  address = {Oxford},
  year = {2003},
  isbn = {9780198525004},
  doi = {10.1093/acprof:oso/9780198525004.001.0001}
}

@article{PhysRevB.92.085138,
  author = {Gos\'albez-Mart\'{\i}nez, Daniel and Souza, Ivo and Vanderbilt, David},
  title = {Chiral degeneracies and {Fermi}-surface {Chern} numbers in bcc {Fe}},
  journal = {Phys. Rev. B},
  publisher = {American Physical Society},
  volume = {92},
  number = {8},
  pages = {085138},
  year = {2015},
  month = aug,
  doi = {10.1103/PhysRevB.92.085138},
  url = {https://doi.org/10.1103/PhysRevB.92.085138},
  numpages = {22}
}

@article{kirschbaum2026emergent,
  author = {Kirschbaum, D. M. and Chen, L. and Zocco, D. A. and Hu, H. and Mazza, F. and Karlich, M. and Lu{\v z}nik, M. and Nguyen, D. H. and {Larrea Jim{\'e}nez}, J. and Strydom, A. M. and Adroja, D. and Yan, X. and Prokofiev, A. and Si, Q. and Paschen, S.},
  title = {Emergent topological semimetal from quantum criticality},
  journal = {Nat. Phys.},
  volume = {22},
  pages = {218--224},
  year = {2026},
  doi = {10.1038/s41567-025-03135-w}
}

@article{dzsaber2022control,
  author = {Dzsaber, Sami and Zocco, Diego A. and McCollam, Alix and Weickert, Franziska and McDonald, Ross and Taupin, Mathieu and Eguchi, Gaku and Yan, Xinlin and Prokofiev, Andrey and Tang, Lucas M. K. and Vlaar, Bryan and Winter, Laurel E. and Jaime, Marcelo and Si, Qimiao and Paschen, Silke},
  title = {Control of electronic topology in a strongly correlated electron system},
  journal = {Nat. Commun.},
  volume = {13},
  pages = {5729},
  year = {2022},
  doi = {10.1038/s41467-022-33369-8}
}

@misc{liu2024antiferromagnetic,
  author = {Liu, D. F. and Xu, Y. F. and Hu, H. Y. and Liu, J. Y. and Ying, T. P. and Lv, Y. Y. and Jiang, Y. and Chen, C. and Yang, Y. H. and Pei, D. and Prabhakaran, D. and Gao, M. H. and Wang, J. J. and Zhang, Q. H. and Meng, F. Q. and Thiagarajan, B. and Polley, C. and Hashimoto, M. and Lu, D. H. and Schr{\"o}ter, N. B. M. and Strocov, V. N. and Louat, A. and Cacho, C. and Biswas, D. and Lee, T.-L. and Steadman, P. and Bencok, P. and Chen, Y. B. and Gu, L. and Hesjedal, T. and van der Laan, G. and Hosono, H. and Yang, L. X. and Liu, Z. K. and Yuan, H. Q. and Bernevig, B. A. and Chen, Y. L.},
  title = {Discovery of an antiferromagnetic topological nodal-line {Kondo} semimetal},
  year = {2024},
  doi = {10.48550/arXiv.2411.13898},
  eprint = {2411.13898},
  archivePrefix = {arXiv},
  primaryClass = {cond-mat.str-el}
}

@article{pirie2020imaging,
  author = {Pirie, Harris and Liu, Yu and Soumyanarayanan, Anjan and Chen, Pengcheng and He, Yang and Yee, M. M. and Rosa, P. F. S. and Thompson, J. D. and Kim, Dae-Jeong and Fisk, Z. and Wang, Xiangfeng and Paglione, Johnpierre and Morr, Dirk K. and Hamidian, M. H. and Hoffman, Jennifer E.},
  title = {Imaging emergent heavy {Dirac} fermions of a topological {Kondo} insulator},
  journal = {Nat. Phys.},
  volume = {16},
  number = {1},
  pages = {52--56},
  year = {2020},
  doi = {10.1038/s41567-019-0700-8}
}

@article{park2016topological,
  author = {Park, Wan Kyu and Sun, Lunan and Noddings, Alexander and Kim, Dae-Jeong and Fisk, Zachary and Greene, Laura H.},
  title = {Topological surface states interacting with bulk excitations in the {Kondo} insulator {SmB$_6$} revealed via planar tunneling spectroscopy},
  journal = {Proc. Natl. Acad. Sci. U.S.A.},
  volume = {113},
  number = {24},
  pages = {6599--6604},
  year = {2016},
  doi = {10.1073/pnas.1606042113}
}

@article{gupta2023topological,
  author = {Gupta, A. and Weiser, A. and Pressley, L. and Luo, Y. and Lygouras, C. and Trowbridge, J. and Phelan, W. A. and Broholm, C. L. and McQueen, T. M. and Park, W. K.},
  title = {Topological surface states in the {Kondo} insulator {YbB$_{12}$} revealed via planar tunneling spectroscopy},
  journal = {Phys. Rev. B},
  volume = {107},
  number = {16},
  pages = {165132},
  year = {2023},
  doi = {10.1103/PhysRevB.107.165132}
}

@article{sato2021surface,
  author = {Sato, Y. and Xiang, Z. and Kasahara, Y. and Kasahara, S. and Chen, Lu and Tinsman, C. and Iga, F. and Singleton, J. and Nair, N. L. and Maksimovic, N. and Analytis, J. G. and Li, Lu and Matsuda, Y.},
  title = {Topological surface conduction in {Kondo} insulator {YbB$_{12}$}},
  journal = {J. Phys. D: Appl. Phys.},
  volume = {54},
  number = {40},
  pages = {404002},
  year = {2021},
  doi = {10.1088/1361-6463/ac10d9}
}

@article{grefe2020nonsymmorphic,
  author = {Grefe, Sarah E. and Lai, Hsin-Hua and Paschen, Silke and Si, Qimiao},
  title = {{Weyl-Kondo} semimetals in nonsymmorphic systems},
  journal = {Phys. Rev. B},
  volume = {101},
  number = {7},
  pages = {075138},
  year = {2020},
  doi = {10.1103/PhysRevB.101.075138}
}

@article{ma2023nodeline,
  author = {Ma, Hao-Tian and Ming, Xing and Zheng, Xiao-Jun and Wen, Jian-Feng and Wang, Yue-Chao and Liu, Yu and Li, Huan},
  title = {Node-line {Dirac} semimetal manipulated via {Kondo} mechanism in nonsymmorphic {CePt$_2$Si$_2$}},
  journal = {Phys. Rev. B},
  volume = {107},
  number = {7},
  pages = {075124},
  year = {2023},
  doi = {10.1103/PhysRevB.107.075124}
}

@article{oshikawa2000luttinger,
  author = {Oshikawa, Masaki},
  title = {Topological approach to {Luttinger's} theorem and the {Fermi} surface of a {Kondo} lattice},
  journal = {Phys. Rev. Lett.},
  volume = {84},
  number = {15},
  pages = {3370--3373},
  year = {2000},
  doi = {10.1103/PhysRevLett.84.3370}
}

@article{dzero2016review,
  author = {Dzero, Maxim and Xia, Jing and Galitski, Victor and Coleman, Piers},
  title = {Topological {Kondo} insulators},
  journal = {Annu. Rev. Condens. Matter Phys.},
  volume = {7},
  pages = {249--280},
  year = {2016},
  doi = {10.1146/annurev-conmatphys-031214-014749}
}

@article{wangDiracSemimetalTopological2012,
  author = {Wang, Zhijun and Sun, Yan and Chen, Xing-Qiu and Franchini, Cesare and Xu, Gang and Weng, Hongming and Dai, Xi and Fang, Zhong},
  title = {Dirac semimetal and topological phase transitions in {$A_3$Bi} ($A = \mathrm{Na}$, {K}, {Rb})},
  journal = {Phys. Rev. B},
  volume = {85},
  number = {19},
  pages = {195320},
  year = {2012},
  doi = {10.1103/PhysRevB.85.195320}
}

@article{wangThreeDimensionalDirac2013,
  author = {Wang, Zhijun and Weng, Hongming and Wu, Quansheng and Dai, Xi and Fang, Zhong},
  title = {Three-dimensional {Dirac} semimetal and quantum transport in {Cd$_3$As$_2$}},
  journal = {Phys. Rev. B},
  volume = {88},
  number = {12},
  pages = {125427},
  year = {2013},
  doi = {10.1103/PhysRevB.88.125427}
}

@misc{SM,  note = {See Supplementary Material for details of the mean-field calculation, the definition of the microscopic model, and the corresponding numerical results. }}

@Article{Xiang2021,
author={Xiang, Ziji
and Chen, Lu
and Chen, Kuan-Wen
and Tinsman, Colin
and Sato, Yuki
and Asaba, Tomoya
and Lu, Helen
and Kasahara, Yuichi
and Jaime, Marcelo
and Balakirev, Fedor
and Iga, Fumitoshi
and Matsuda, Yuji
and Singleton, John
and Li, Lu},
title={Unusual high-field metal in a Kondo insulator},
journal={Nature Physics},
year={2021},
month={Jul},
day={01},
volume={17},
number={7},
pages={788-793},
issn={1745-2481},
doi={10.1038/s41567-021-01216-0},
url={https://doi.org/10.1038/s41567-021-01216-0}
}

@article{PhysRevB.105.104515,
  title = {Euler-obstructed Cooper pairing: Nodal superconductivity and hinge Majorana zero modes},
  author = {Yu, Jiabin and Chen, Yu-An and Das Sarma, Sankar},
  journal = {Phys. Rev. B},
  volume = {105},
  issue = {10},
  pages = {104515},
  numpages = {40},
  year = {2022},
  month = {Mar},
  publisher = {American Physical Society},
  doi = {10.1103/PhysRevB.105.104515},
  url = {https://link.aps.org/doi/10.1103/PhysRevB.105.104515}
}

\onecolumngrid
\begin{center}
\textbf{Appendix}\\ 
\end{center}

\renewcommand{\thefigure}{S\arabic{figure}}

\renewcommand{\thetable}{S\arabic{table}}

\renewcommand{\thesection}{S\arabic{section}}

\renewcommand{\theequation}{S\arabic{equation}}
\setcounter{equation}{0}
\setcounter{figure}{0}
\renewcommand{\theHfigure}{S\arabic{figure}}
\setcounter{table}{0}
\setcounter{section}{0}

\tableofcontents

\clearpage

\section{Kondo lattice model and Kondo effect} 
We briefly review the derivation of the Kondo lattice model from the Anderson lattice model via the Schrieffer--Wolff transformation, as well as the saddle-point solution of the Kondo lattice model. 

We start from a generic Anderson lattice model described by the Hamiltonian
\begin{align}
H_{AL} = &
\sum_{\kk,i,j,\sigma,\sigma'} t_{\kk,i\sigma,j\sigma'}c_{\kk,i\sigma}^\dag c_{\kk,j\sigma'}  + \frac{1}{\sqrt{N}} \sum_{\kk,\RR,\sigma,\sigma',m} \bigg[ [W_{\kk}]_{m\sigma',\sigma}^* f_{\RR,\sigma}^\dag c_{\kk,m\sigma'} e^{i\kk\cdot(\RR+\rr_f)}
+\text{h.c.}\bigg] \nonumber\\ 
& +\frac{U}{2} \sum_{\RR} \bigg(\sum_\sigma f_{\RR,\sigma}^\dag f_{\RR,\sigma}-1\bigg)^2 .
\end{align}
Here $c_{\kk,i\sigma}^\dag$ creates a conduction electron with momentum $\kk$, orbital index $i$, and spin $\sigma$, located at position $\rr_i$. The operator $f_{\RR,\sigma}^\dag$ creates an $f$ electron with spin $\sigma$ in the unit cell at $\RR$, located at $\rr_f$. To simplify the analysis, we consider only a single $f$ orbital per unit cell. 
The matrix $t_{\kk,i\sigma,j\sigma'}$ describes the hopping of conduction electrons, while $U$ denotes the on-site Hubbard interaction of the $f$ electrons. The matrix $W_{\kk}$ is defined by $H_{\mathrm{hyb}}=\sum_\kk(c_\kk^\dagger W_\kk f_\kk+\mathrm{h.c.})$. We use the Fourier convention $c_{\kk,i\sigma}=N^{-1/2}\sum_{\RR} e^{-i\kk\cdot(\RR+\rr_i)}c_{\RR,i\sigma}$, and analogously for $f$ with $\rr_i$ replaced by $\rr_f$.

We focus on a system with $\mathcal{PT}$ symmetry, which ensures twofold degeneracy of the conduction-electron bands. 
We now introduce the conduction-electron band basis, 
\begin{align}
   \gamma_{\kk,ns}^\dag = \sum_{i\sigma} U_{\kk, i\sigma, ns}c_{\kk,i,\sigma}^\dag 
\end{align}
where $U_{\kk,i\sigma,n s}$ is the eigenvector of the hopping matrix with eigenvalue $\epsilon_{\kk,n}$ 
\begin{align} 
\sum_{j\sigma'}t_{\kk,i\sigma,j\sigma'} U_{\kk,j\sigma',ns}
=\epsilon_{\kk,n} U_{\kk,i\sigma,ns}
\end{align} 
The states labeled by $s=+$ and $s=-$ are related by $\mathcal{PT}$ symmetry and have the same energy. 
In this basis, the Hamiltonian becomes 
\begin{align} 
H_{AL} = &
\sum_{\kk,ns}\epsilon_{\kk,n} \gamma_{\kk,ns}^\dag \gamma_{\kk,ns}  + \frac{1}{\sqrt{N}} \sum_{\kk,\RR,\sigma,s,n} \bigg[ [V_\kk]_{ns,\sigma}^* 
f_{\RR,\sigma}^\dag \gamma_{\kk,ns} e^{i\kk\cdot(\RR+\rr_f)}
+\text{h.c.}\bigg] \nonumber\\ 
& +\frac{U}{2} \sum_{\RR} \bigg(\sum_\sigma f_{\RR,\sigma}^\dag f_{\RR,\sigma}-1\bigg)^2 .
\end{align} 
where the hybridization in the band basis is defined as 
\begin{align} 
[V_{\kk}]_{ns,\sigma} = \sum_{m,\sigma'}
U_{\kk,m\sigma',ns}^*[W_{\kk}]_{m\sigma',\sigma},\qquad V_\kk=U_\kk^\dagger W_\kk .
\end{align}

In the local-moment regime, with conduction-electron energies small compared with the charge-excitation energy $U/2$, charge fluctuations of the $f$ electrons are suppressed and the $f$ degrees of freedom behave as local moments. In this regime, the Anderson lattice model can be mapped to the Kondo lattice model via a Schrieffer--Wolff transformation,
\begin{align} 
    H_{\text{KL}} = \sum_{\kk,ns}\epsilon_{\kk,n} \gamma_{\kk,ns}^\dag \gamma_{\kk,ns}
    + \sum_{\substack{
    \kk,\kk',\RR \\
    m,m' \\
    \sigma, s, \sigma', s'
    }}
    \frac{ {J}_{\kk,\kk'}^{\sigma m s, {\sigma'} m' s' } }{N} 
    f_{\RR,\sigma}^\dag f_{\RR,\sigma'} \gamma^{\dag}_{\kk',m' s'} \gamma_{\kk, m s}  e^{i(\kk-\kk') \cdot (\RR + \rr_f)}
\end{align} 
where we have omitted the single-particle term, which can be absorbed into the kinetic energy of the conduction electrons. The Kondo coupling is 
\begin{align} 
J_{\kk,\kk'}^{\sigma ms,\sigma'm's'}
= \frac{4}{U}[V_{\kk}]_{ms,\sigma}^*[V_{\kk'}]_{m's',\sigma'}
\end{align} 
For a scalar hybridization $v$, this yields $J_K\mathbf S_f\cdot\mathbf s_c$ with $\mathbf s_c=c^\dagger\boldsymbol\sigma c/2$. The independent potential-scattering term vanishes at the symmetric $f$ level used here and is omitted, together with an additive constant.
The $f$ electrons additionally satisfy the local constraint
\begin{align}
\sum_\sigma f_{\RR,\sigma}^\dag f_{\RR,\sigma}=1 .
\end{align}

To decouple the exchange, define
\begin{align}
 \eta_{\RR,\sigma}&=\frac{1}{\sqrt N}\sum_{\kk,ns}e^{i\kk\cdot(\RR+\rr_f)}[V_\kk]_{ns,\sigma}^*\gamma_{\kk,ns},\\
 B_{\RR}&=\sum_\sigma f_{\RR,\sigma}^\dagger\eta_{\RR,\sigma}.
\end{align}
Fermionic normal ordering gives
\begin{align}
 H_K=-\frac4U\sum_{\RR} B_{\RR}^\dagger B_{\RR}
 +\frac2U\sum_{\RR,\sigma}\eta_{\RR,\sigma}^\dagger\eta_{\RR,\sigma}.
\end{align}
We absorb this one-body contribution into the conduction Hamiltonian, $h_c(\kk)=t_\kk+(2/U)W_\kk W_\kk^\dagger$. Below, $\epsilon_{\kk,n}$ and $V_\kk=U_\kk^\dagger W_\kk$ refer to the eigenvalues and hybridization amplitudes in the band basis of this renormalized Hamiltonian.

We analyze the Kondo lattice model within the saddle-point approximation. To this end, we introduce the mean-field amplitude 
\begin{align}
\chi =\frac{4}{U} \frac{1}{\sqrt{N}}\sum_\kk \sum_m \sum_{\sigma,s}\langle f_{\RR,\sigma}^\dag \gamma_{\kk,m s} \rangle e^{i\kk\cdot(\RR+\rr_f)}[V_{\kk}]_{ms,\sigma}^* .
\end{align}
Applying a mean-field decoupling leads to the Hamiltonian
\begin{align}
H_{KL}^{MF} =  &\sum_{\kk,ns}\epsilon_{\kk,n} \gamma_{\kk,ns}^\dag \gamma_{\kk,ns} -  \sum_{\kk,n,\sigma,s} \bigg[ [V_{\kk}]_{ns,\sigma}  \chi
\gamma^\dag_{\kk,ns }  f_{\kk,\sigma} +\text{h.c.}\bigg]
	\nonumber\\
    &+\frac{U}{4}N|\chi|^2  + \lambda_f \bigg(\sum_{\kk,\sigma} f_{\kk,\sigma}^\dag f_{\kk,\sigma}-N\bigg) .
\end{align}
Here $\lambda_f$ is the Lagrange multiplier enforcing the local occupancy constraint of the $f$ electrons.

The effective mean-field Hamiltonian can therefore be written as 
\begin{align} 
H_{KL}^{MF} = 
-N \lambda_f 
+\frac{U}{4}N|\chi|^2 
+ \sum_{\kk}
\bigg[ 
\sum_{n,s} \epsilon_{\kk,n}\gamma_{\kk,ns}^\dag \gamma_{\kk,ns} 
+\sum_{\sigma} \lambda_f f_{\kk,\sigma}^\dag f_{\kk,\sigma} 
- \sum_{ns\sigma}\bigg( [\Delta_{\kk,n}]_{s\sigma} \gamma_{\kk,n s}^\dag f_{\kk,\sigma}
+\text{h.c.}\bigg) 
\bigg] 
\end{align} 
where the hybridization field reads 
\begin{align}
    [\Delta_{\kk,n}]_{s\sigma}
    =\chi [V_{\kk}]_{ns,\sigma} 
\end{align}
One can solve the mean-field Hamiltonian and determine the order parameter $\chi$ self-consistently, together with the hybridization matrix $\Delta_{\kk,ns,\sigma}$.

Near the Fermi energy, when the $n$th band forms a Fermi surface, the low-energy degrees of freedom are described by the conduction-electron operators $\gamma_{\kk,ns}$ and the $f$-electron operators $f_{\kk,\sigma}$. Restricting to these states leads to the following effective Hamiltonian:
\begin{align} 
   & \psi_{\kk,n}^\dag  
    \begin{bmatrix}
    \epsilon_{\kk,n} \mathbb{I}_{2\times 2} & -\Delta_{\kk,n} \\
    -\Delta^\dag_{\kk,n} & \lambda_f \mathbb{I}_{2\times 2}
    \end{bmatrix} \psi_{\kk,n}\nonumber\\
    & \psi_{\kk,n} = \begin{bmatrix}
    \gamma_{\kk,n,+}&\gamma_{\kk,n,-}& f_{\kk,\uparrow}&f_{\kk,\downarrow}
    \end{bmatrix} ^T 
\end{align}   
Diagonalizing this Hamiltonian gives the dispersion in the Kondo phase,
\begin{align} 
E_{\kk,n,\pm}\approx 
\frac{\epsilon_{\kk,n}+\lambda_f}{2} 
\pm \sqrt{ 
\frac{(\epsilon_{\kk,n}-\lambda_f)^2}{4} 
+\Lambda_{\kk,n}^2 
}
\end{align} 
where $\Lambda_{\kk,n}$ denotes the common singular value of the hybridization matrix $\Delta_{\kk,n}$ under the symmetry assumptions stated in the main text.


\section{Microscopic model}
\label{sec:tr_model}
We consider a microscopic model with $D_{4h}$ symmetry. The model contains two conduction-electron Kramers doublets, labeled $s$ and $p$, centered at $\rr_c=(0,0,0)$ and transforming under the irreducible representations $\Gamma_6^+$ and $\Gamma_7^-$, respectively. A localized $f$-electron Kramers doublet is centered at $\rr_f=(1/2,1/2,0)$ and transforms under $\Gamma_7^-$. 
 In the basis $c_{\kk,\sigma}=(c_{\kk,s,\sigma},c_{\kk,p,\sigma})^T$, the conduction Hamiltonian is
\begin{align}
 H_c&=\sum_{\kk,\sigma}c_{\kk,\sigma}^{\dagger}h_\sigma(\kk)c_{\kk,\sigma},\label{eq:model_Hc}\\
 h_\up(\kk)&=-\mu\tau_0+t_x\sin k_x\,\tau_1+t_y\sin k_y\,\tau_2+M(\kk)\tau_3,\label{eq:model_hc}\\
 h_\dn(\kk)&=h_\up^*(-\kk),\label{eq:model_hdown}\\
 M(\kk)&=m+t_z\cos k_z+2b(2-\cos k_x-\cos k_y).\label{eq:model_mass}
\end{align}
Here $\tau_i$ act in orbital space. The model preserves time-reversal symmetry and $U(1)$ spin symmetry. The two spin sectors share the conduction-band dispersions
\begin{equation}
 \epsilon_{\kk,\pm}=-\mu\pm\sqrt{t_x^2\sin^2k_x+t_y^2\sin^2k_y+M(\kk)^2}.
 \label{eq:model_c_dispersion}
\end{equation}

\begin{figure}[!htbp]
 \centering
 \includegraphics[width=0.5\linewidth]{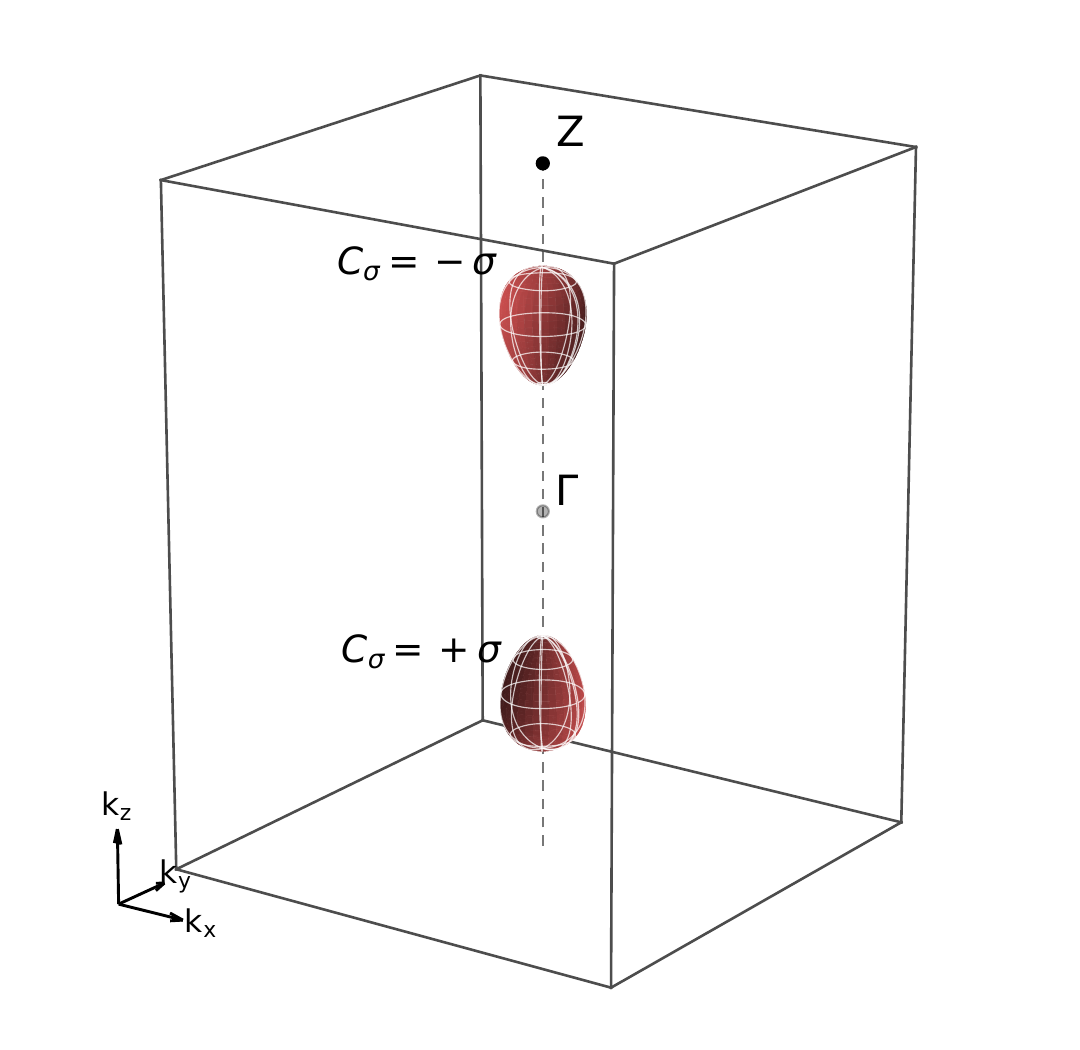}
 \caption{Fermi surface of the conduction electrons. We use $t_x=t_y=t_z=t$, $m/t=0.1$, $b/t=0.5$, and $\mu/t=0.5$.}
 \label{fig:model_fermi_surface}
\end{figure}

We use $t_x=t_y=t_z=t$, $m/t=0.1$, $b/t=0.5$, and $\mu/t=0.5$. As shown in Fig. \ref{fig:model_fermi_surface}, the Fermi surface of conduction electrons consists of two separate pockets surrounding the Dirac points $\kk_\pm=(0,0,\pm k_0)$, where $k_0=\arccos(-m/t_z)$. For each Fermi pocket, the two spin sectors carry opposite nonzero Chern numbers. 
To introduce the Kondo coupling, we start from the $fc$ hybridization
\begin{align}
 H_{\mathrm{hyb}}&=\sum_{\kk,a,\sigma}\left[c_{\kk,a,\sigma}^\dagger W_a^\sigma(\kk)f_{\kk,\sigma}+\mathrm{h.c.}\right],\\
 W^\sigma(\kk)&=\begin{pmatrix}
 4V_s\left[\sigma\sin(k_x/2)\cos(k_y/2)+i\cos(k_x/2)\sin(k_y/2)\right]\\
 4V_p\cos(k_x/2)\cos(k_y/2)
 \end{pmatrix}.
 \label{eq:model_W}
\end{align}
Here $\sigma=\pm1$. The real parameters $V_s$ and $V_p$ set the hybridization strengths with the $s$ and $p$ orbitals, respectively. 
The corresponding Kondo interaction takes the form
\begin{align}
 H_K&=\frac{1}{N}\sum_{\RR,\kk,\kk'}\sum_{a,b,\sigma,\sigma'}
 J_{ab}^{\sigma\sigma'}(\kk,\kk')e^{i(\kk-\kk')\cdot(\RR+\rr_f)}
f_{\RR,\sigma}^\dagger f_{\RR,\sigma'}
 c_{\kk',b,\sigma'}^\dagger c_{\kk,a,\sigma},\\
 J_{ab}^{\sigma\sigma'}(\kk,\kk')&=\frac{4}{U}\left[W_a^\sigma(\kk)\right]^*W_b^{\sigma'}(\kk'),\qquad a,b\in\{s,p\},
 \label{eq:model_exchange}
\end{align}

\begin{figure}[!htbp]
 \centering
 \includegraphics[width=0.98\linewidth]{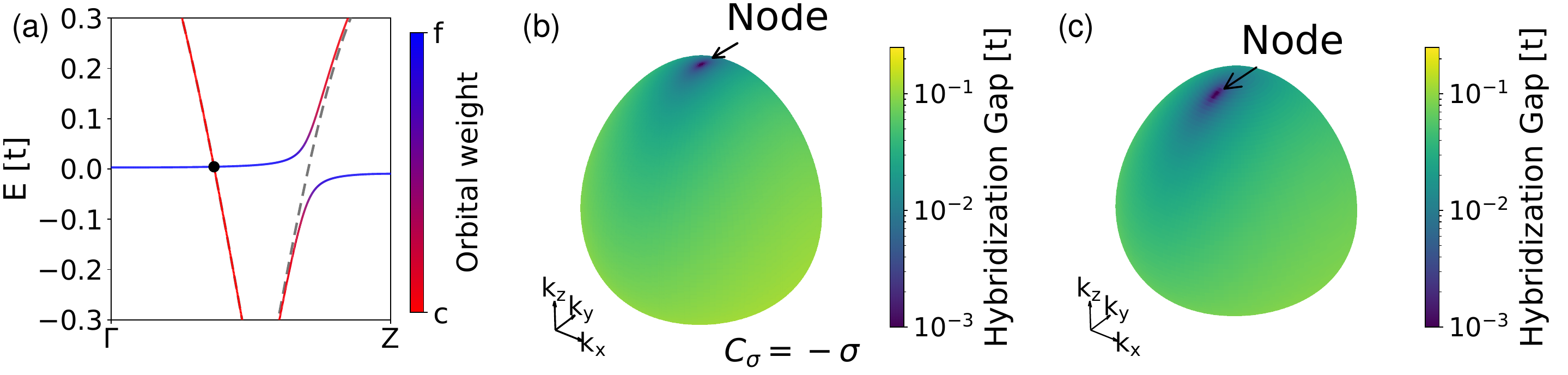}
 \caption{Kondo dispersion and hybridization gap. (a) Dispersion along the path from $\Gamma=(0,0,0)$ to $Z=(0,0,\pi)$. The colors of the solid lines indicate the orbital weights: red indicates conduction character, and blue indicates $f$ character. Gray dashed lines show the conduction-electron dispersion in the absence of the Kondo effect, and the black dot marks the hybridization node. (b) Kondo hybridization gap $2|\Delta_{\kk,n,\sigma}|/t$ on a conduction-electron Fermi pocket. (c) Hybridization gap after introducing the symmetry-breaking perturbation with $\theta=0.1\pi$, $\delta=0.1$, and $\eta=0.05t$. In (b) and (c), the logarithmic color scale is truncated below $10^{-3}$, and the arrow marks the node. We use $t_x=t_y=t_z=t$, $m/t=0.1$, $b/t=0.5$, $\mu/t=0.5$, $V_s=0.35t$, $V_p=0.5t$, $U=10t$, and $\beta t=10^4$.}
 \label{fig:model_tr_dispersion}
\end{figure}

We further set $V_s=0.35t$, $V_p=0.5t$, $U=10t$, and $\beta t=10^4$. A self-consistent mean-field calculation on a $24^3$ momentum grid yields a Kondo-screened state. As shown in \cref{fig:model_tr_dispersion}(a), the Kondo bands retain a node near the chemical potential, demonstrating the obstruction to opening a full hybridization gap. 
To visualize this obstruction, we evaluate the size of the hybridization gap on the effective conduction-electron Fermi surface ($\epsilon_{\kk,+}=\lambda_f$). \Cref{fig:model_tr_dispersion}(b) shows the result for one of the Fermi pockets. The hybridization gap is finite everywhere except at a single point, revealing the node required by the topological properties of the Fermi surface. 

To show the robustness of the node, we generalize the upper component of \cref{eq:model_W} to 
\begin{equation}
[ W^\sigma(\kk)]_1 = 4V_s e^{i\sigma\theta}\left[\sigma\sin\frac{k_x}{2}\cos\frac{k_y}{2}+i(1-\delta)\cos\frac{k_x}{2}\sin\frac{k_y}{2}\right]
+4\left(i\eta\sin k_z\right)\cos\frac{k_x}{2}\cos\frac{k_y}{2}.
\label{eq:app_W_chain}
\end{equation}
Nonzero values of $\theta$, $\delta$, and $\eta$ break rotational and mirror symmetries of the system. We perform the self-consistent mean-field calculation with $\theta=0.1\pi$, $\delta=0.1$, and $\eta=0.05t$. The results in \cref{fig:model_tr_dispersion}(c) again show a node in the Kondo phase.

\end{document}